\documentclass[aps,prd,twocolumn,nofootinbib,amsmath,amssymb,floatfix]{revtex4}
\usepackage{graphicx}  
\usepackage{dcolumn}   
\usepackage{bm,color}        
\usepackage{amssymb}   
\usepackage{blkarray}   

\renewcommand{\vec}{\bm}

\newcommand{\be}{\begin{equation}}
\newcommand{\ee}{\end{equation}}

\newcommand{\ihMpc}{h\text{Mpc}^{-1}}
\newcommand{\knl}{k_{\rm NL}}
\newcommand{\kb}{\vec k}

\newcommand{\comment}[1]{}

\usepackage{soul}
\usepackage{enumitem}

\definecolor{myBlue}{RGB}{15, 40, 220}
\usepackage[linktocpage=true]{hyperref}
\hypersetup{colorlinks=true, citecolor=myBlue, linkcolor=myBlue, urlcolor=violet}

\begin{document}

\title{LSS constraints with controlled theoretical uncertainties}

\author{Tobias Baldauf}
\affiliation{Institute for Advanced Study, Einstein Drive, Princeton, NJ 08540, USA}
\author{Mehrdad Mirbabayi}
\affiliation{Institute for Advanced Study, Einstein Drive, Princeton, NJ 08540, USA}
\author{Marko Simonovi\'c}
\affiliation{Institute for Advanced Study, Einstein Drive, Princeton, NJ 08540, USA}
\author{Matias Zaldarriaga}
\affiliation{Institute for Advanced Study, Einstein Drive, Princeton, NJ 08540, USA}


\begin{abstract}
Forecasts and analyses of cosmological observations often rely on the assumption of a perfect theoretical model over a defined range of scales. We explore how model uncertainties and nuisance parameters in perturbative models of the matter and galaxy spectra affect constraints on neutrino mass and primordial non-Gaussianities. We provide a consistent treatment of theoretical errors and argue that their inclusion is a necessary step to obtain realistic cosmological constraints. We find that galaxy surveys up to high redshifts will allow a detection of the minimal neutrino mass and local non-Gaussianity of order unity, but improving the constraints on equilateral non-Gaussianity beyond the CMB limits will be challenging. We argue that similar considerations apply to analyses where theoretical models are based on simulations.
\end{abstract}

\maketitle

\section{Introduction}
\noindent
We live in the era of precision cosmology, mainly due to the observations of temperature and polarization fluctuations in the Cosmic Microwave Background (CMB). All six parameters of the minimal $\Lambda$CDM model are measured with high accuracy \cite{Ade:2015xua}. However, for many interesting quantities only upper limits are known. In some examples, like the sum of neutrino masses, further improvements are possible from the CMB alone. In some others, like primordial non-Gaussianities (NG), the upper limits are close to their cosmic variance limited values. In such cases, in order to make progress, one has to find a way to observe more independent modes elsewhere.

Luckily, a lot of additional information is available in the Large Scale Structure (LSS) of the Universe. While the number of modes in the CMB is approximately $N_{\rm CMB}\sim l_{\rm max}^2 \sim 10^7$, in the LSS it scales like $N_{\rm LSS}\sim (k_{\rm max}/k_{\rm min})^3$ and can be much bigger. The signal-to-noise is proportional to $\sqrt{N_{\rm LSS}}$ and it grows with the volume of a survey and $k_{\rm max}$. While mapping larger volumes is mainly an observational challenge, increasing $k_{\rm max}$ at which we can reliably compare theories and observations is a theoretical problem. High values of $k_{\rm max}$ correspond to short scales in real space, where we expect gravitational nonlinearities in the density fluctuations to become important. Our theoretical description of these mildly nonlinear scales -- whether analytical or numerical -- is not perfect and observables are calculated with finite theoretical precision. The theoretical uncertainty, which can be thought of as a systematic error, is often neglected but can have very important consequences. Simply put, if the signal of the new physics is within the theoretical uncertainties, we are not guaranteed to be able to detect it even if the other (observational) systematic errors are very small. The era of high precision data requires equally precise theoretical predictions. 

The available precision of theoretical calculations depends on the approach. One example is the analytical treatment of nonlinearities in which one solves the equations of motion perturbatively to find the nonlinear density contrast $\delta$ (for a review see \cite{Bernardeau:2001qr}). The perturbation theory approach was recently put on solid footing in the framework of the Effective Field Theory (EFT) of LSS \cite{Baumann:2010tm,Carrasco:2012cv,Carrasco:2013mua,Hertzberg:2012qn}, which provides both a way to treat the back-reaction of small scale nonlinearities on larger scales, and an estimate of the size of the subleading corrections. In a toy model of a scaling universe with linear power spectrum $P(k)\propto k^n$, higher order corrections to the power spectrum have a simple form. At a given number of loops $l$, the relative contribution with respect to the leading tree-level result is of order $(k/k_{\rm NL})^{(3+n)l}$ \cite{Pajer:2013jj}. Here, $\knl$ is the wavenumber below which the perturbation theory makes sense.\footnote{The real universe is not a scaling universe, but for the mildly nonlinear range of modes, scaling is a good approximation with $n\approx -1.5$ and $k_{\rm NL}\approx 0.3\; h/{\rm Mpc}$ at redshift zero (see for example \cite{Baldauf:2015tla}).} For $k<k_{\rm NL}$ perturbation theory with a finite number of loops is still just an approximation to the true answer. The characteristic size of the corrections can be estimated by the next loop order which decreases as the number of loops increases, and it is always smaller further below the nonlinear scale. 

The other strategy for dealing with the nonlinearities is to solve the equations of motion numerically. This is achieved in $N$-body simulations. Simulations with dark matter only in principle give the correct answer even in the nonlinear regime. In practice, due to complexity of the problem, many approximations have to be made and this again leads to errors. These errors depend on the details of algorithms and they are hard to estimate, but the typical precision that the current simulations can achieve is $\mathcal O(1\%)$ with a mild scale dependence (see for example \cite{Heitmann:2008eq,Schneider:2015yka}). The simulations that include the effects of baryonic physics are even less reliable. They do not solve first principle equations of motion and they contain many free parameters which can be degenerate with the signal of interest.

In this paper we explore how much these theoretical errors, present both in perturbative approaches and simulations, affect forecasts for LSS surveys and CMB lensing. This should be contrasted with the usual assumption of perfect knowledge of relevant quantities all the way up to some arbitrarily chosen scale $k_{\rm NL}$. Theoretical errors effectively restrict the range of useful modes to those for which the signal dominates over the theoretical uncertainty. In this way the realistic $k_{\rm max}$ can be surprisingly lower than $k_{\rm NL}$ and this reduction of the number modes leads to bigger uncertainties on inferred parameters.

We will describe in detail how to consistently calculate the Fisher matrix including the theoretical uncertainties. We will apply this general framework to measurements of the sum of neutrino masses and primordial NG (for a similar earlier study for the case of neutrino mass see \cite{Audren:2012vy}). Obtaining realistic and very precise forecasts, particularly for very high redshift surveys, is beyond the scope of this paper. Our primary goal is to study the effect of theoretical uncertainties on the amount of useful information in a given volume. We will therefore use simple analytical models whenever possible and assume ideal surveys. In this sense our final results are optimistic, but nevertheless give a very good estimate of how much theoretical errors degrade the constraints. 

Before moving to the more systematic treatment, in the rest of this section we motivate the basic idea in the example of equilateral NG.

\subsection{Example of Equilateral NG}
\noindent
Primordial NG are important observables because they contain information about the very early phases of cosmic evolution. The current upper bounds on the most interesting equilateral and local shapes are \cite{Ade:2015ava}
\be
f_{\rm NL}^{\rm loc.}=0.8\pm5.0 \;, \quad f_{\rm NL}^{\rm eq.}=-4\pm43\;, \quad (68\%\; {\rm CL}) \;.
\ee
Even though these upper limits are quite strong, a theoretically interesting threshold is $f_{\rm NL}\sim1$. Any detection of non-zero NG would be very exciting, but even the observation that both $f_{\rm NL}^{\rm loc.}$ and $f_{\rm NL}^{\rm eq.}$ are smaller than one would be very informative. It would favor single-field and slow-roll inflation and practically rule out a large class of inflationary models with modified kinetic term or more than one light field during inflation. Although futuristic experiments including polarization have a potential to improve the current constraints almost by a factor of 2 (see for example \cite{Koyama:2013wma}), it will be hard to reach $f_{\rm NL}\sim1$ from the CMB alone.

The other way to detect primordial NG is through its imprint on the bispectrum of density fluctuations in the late universe. The full bispectrum $B({\kb}_1,{\kb}_2,{\kb}_3)$ of the density contrast $\delta$ is a sum of the primordial part and the one generated by the gravitational interactions. For simplicity, let us focus on redshift $z=0$ and assume that all momenta in the bispectrum are of the same magnitude $k$. The primordial contribution is approximately
\be
B^{\rm eq.}(k) \sim P^2(k) \cdot f_{\rm NL}^{\rm eq.} \frac{9H_0^2\Omega_{\rm m}}{k^2T(k)D_+(0)} \;,
\ee
where $T(k)$ is the transfer function, $H_0$ the present day value of the Hubble constant, $\Omega_{\rm m}$ the matter density parameter and $D_+(z)$ the perturbation growth factor. The gravitational part can be calculated using perturbation theory. If one calculates the bispectrum including $(l-1)$ loops, the result can be schematically written as
\be
\label{B0_grav}
B^{\rm grav.}(k) \sim P^2(k) \left[``(l-1){\rm -loop}" + E(l,k) \right]\;,
\ee
where the second term is the theoretical error. As we discussed, the typical size of this error is $E(l,k)=\mathcal O((k/k_{\rm NL})^{(3+n)l})$. Notice that for the leading tree-level bispectrum the first term in square brackets is $\mathcal O(1)$. 

From the previous expressions it is clear that while the theoretical error grows, the primordial part decays with $k$. We are interested in the scale $k_{\rm max}$ for which they become comparable. This scale sets the range of modes that we are allowed to use in the analysis:
\be
\label{eq:kmax_def}
f_{\rm NL}^{\rm eq.} \frac{9H_0^2\Omega_m}{k_{\rm max}^2T(k_{\rm max})D_+(0)} \sim \left( \frac{k_{\rm max}}{k_{\rm NL}} \right)^{(3+n)l} \;.
\ee
For example, if we calculate the 1-loop bispectrum (corresponding to $l=2$ for the error), for a target of $f_{\rm NL}^{\rm eq.}\sim1$ it turns out that $k_{\rm max}=0.03\; h{\rm Mpc}^{-1}$. This is quite smaller than the naive cutoff $k_{\rm NL}$ and deep in the perturbative regime. On second thought, this result should not be so surprising. For the given $k_{\rm max}$ and $f_{\rm NL}^{\rm eq.} \sim 1$ the relative size of primordial part is
\be
f_{\rm NL}^{\rm eq.} \frac{9H_0^2\Omega_m}{k_{\rm max}^2T(k_{\rm max})D_+(0)} \sim \mathcal O(10^{-3}) \;,
\ee
which should be compared with the $\mathcal O(1)$ gravitational contribution in Eq.~\eqref{B0_grav}. To get this precision on the gravitational bispectrum one has to stay far away from the nonlinear scale. This precision is an order of magnitude smaller than the usual theoretical target, which is $\mathcal O(1\%)$. This is true for perturbation theory as well as for simulations. In order to be useful for detection of small equilateral NG, the theoretical models have to significantly improve.  

So far we were just comparing primordial and gravitational signal to estimate $k_{\rm max}$. It is interesting to ask whether $f_{\rm NL}^{\rm eq.} \sim 1$ is even achievable with $k_{\rm max}=0.03\; h/{\rm Mpc}^{-1}$ and what kind of survey volume is needed. To find the answer we have to calculate the signal-to-noise, which is given by
\be
\begin{split}
\left(\frac SN \right)^2 &= \frac{V^2}{(2\pi)^6} \int d^3k_1d^3k_2d^3k_3 \frac{B^{\rm eq.}({\kb}_1,{\kb}_2,{\kb}_3)^2}{P(k_1)P(k_2)P(k_3)}\\
&\approx \frac{V}{(2\pi)^3} k_{\rm max}^3 {f_{\rm NL}^{\rm eq.}}^2 \mathcal{A} \cdot \mathcal O(1)\;,
\end{split}
\ee
where $\mathcal{A}=2.215\cdot 10^{-9}$ is the normalization of the power spectrum. This can be rewritten as $\sigma(f_{\rm NL}^{\rm eq.}) \sim 2\cdot10^4/\sqrt{N}$, where $N=(k_{\rm max}/k_{\rm min})^3$ is the number of modes. With NG of order unity we naively get $k_{\rm min}\sim10^{-3}k_{\rm max}$ which, for the above estimate of $k_{\rm max}$, corresponds to unobservable super-horizon scales.

The moral of this simple exercise is that reaching $f_{\rm NL}^{\rm eq.}\sim 1$ in future galaxy surveys seems unrealistic. This is not due to the lack of information in LSS, but due to our inability to model the bispectrum more accurately. Obviously, a more careful analysis should be done taking into account many effects neglected in this simplified picture. The most important one is redshift dependence of all relevant quantities. Naively, the main improvement in the constraints when going to higher redshifts is expected to come from the increase of $k_{\rm NL}$. For spectral index $n= -1.5$ the time dependence of the nonlinear scale is $k_{\rm NL}(z) \sim D_+^{-4/3}(z)$. However, $k_{\rm max}$ does not necessarily grow as fast as $k_{\rm NL}$. Using a more detailed signal to noise estimate in section \ref{sec:results}, which includes shot noise and marginalization over bias and EFT parameters, we will tentatively argue that even with futuristic high redshift galaxy surveys a detection of $f_{\rm NL}^{\rm eq.}\sim 10$ will be challenging. This is well above $f_{\rm NL}^{\rm eq.}\sim 1$, which is an interesting threshold for non-trivial dynamics beyond the slow-roll regime \cite{Baumann:2014cja}.

{\section{The forecasting method}
\noindent
In this section we present a method to systematically implement theoretical errors in the forecasts. We first briefly review the standard analysis. The starting point is the Gaussian likelihood given by
\be
\mathcal L =\frac{1}{\sqrt{(2\pi)^{N_\text{c}} |C_d| }}\exp\left[-\frac{1}{2}(\vec d-\vec t) C_d^{-1} (\vec d-\vec t) \right]\; ,
\ee
where $N_\text{c}$ is a number of different momentum configurations that contribute to the likelihood,\footnote{In the case of the power spectrum $N_\text{c}$ is equal to the number of bins, and for the bispectrum to the number of different triangles.} $\vec d$ is a vector of data points at these configurations, $\vec t$ is a vector of theoretical predictions that depend on a number of relevant cosmological parameters and $C_d$ is a covariance matrix $(C_d)_{ij}\equiv\langle \vec d_i\vec d_j \rangle$.

Once the likelihood is known as a function of cosmological parameters, one can calculate the Fisher matrix
\be
\label{eq:fisher_definition}
F_{ij} = -\left \langle \frac{\partial^2\log \mathcal L}{\partial p_i \partial p_j} \right \rangle\Big|_{\vec p = \vec p_0} \;,
\ee
where $\vec p$ is a set of relevant parameters and $\vec p_0$ a set of their fiducial values. The Fisher matrix contains information about how well each of the parameters can be constrained. If one is interested in just one of them, and marginalize over all others, the minimal variance is given by
\be
\sigma(p_i) =\sqrt{(F^{-1})_{ii}} \;.
\ee
The unmarginalized error on one single parameter is given by the inverse of the respective element of the Fisher matrix
\be
\sigma(p_i) =1/\sqrt{F_{ii}} \;.
\ee

\subsection{Including theoretical errors}
\noindent
The most straightforward way to include the theoretical error in the likelihood is to model its shape with a certain finite set of test functions $\vec g_i$ and their associated coefficients $c_i$ and to add this template $\sum c_i \vec g_i$ to the theoretical prediction. Then one can proceed in the usual way, marginalizing over the nuisance parameters $c_i$. 
We found it more convenient to employ a different strategy: The theoretical error $\vec e$ is the difference between the true theory $\vec t_\text{t}$ and the explicitly calculated, fiducial theoretical prediction $\vec t_\text{f}$. The error is bounded by an envelope $\vec E$, but it cannot have arbitrarily fast variations as a function of wavenumber. The characteristic scale of variations of the error is a physical input, which we choose to be the scale of the Baryon Acoustic Oscillations (BAOs) $\Delta k=\Delta k_\text{BAO} = 0.05 \;h{\rm Mpc}^{-1}$, because this is the smallest typical scale over which power spectra vary.\footnote{Baryons are only a small fraction of the total mass, so one might expect that the wiggle part of the power spectrum with $\Delta k =\Delta k_\text{BAO}$ is small compared to a smooth contribution with much larger coherence length. In principle, one can treat this situation with two independent theoretical errors with different $\Delta k$. In practice we do not do it for two reasons: (a) The explicit calculation shows that at scales of interest smooth one-loop and two-loop power spectra cross zero at several points, at which the wiggle contribution dominates and the coherence length of the total power spectrum is indeed $\Delta k_\text{BAO}$. (b) Most of the signal in all our forecasts comes from a range of $k$ which spans only a few coherence lengths. Therefore, we expect that choosing larger $\Delta k$ does not change the results significantly.}

Therefore, we allow for one free parameter in each momentum configuration (wavenumber bin for the power spectrum) and add it to the model. The range of these free parameters for each configuration is determined by the envelope $\vec E$. In perturbative treatments $\vec E$ is of the order of the first neglected loop result and can be estimated in the EFT. For simplicity, we assume that each free parameter has a Gaussian distribution with zero mean and variance $\vec E$. The characteristic scale of variation $\Delta k$ can then be implemented as a correlation between the errors of close momentum configurations, making the covariance of the Gaussian off-diagonal. Finally, we will marginalize over the error. 

Including the error $\vec e$ and its Gaussian prior in the likelihood is straightforward
\begin{equation}
\label{eq:starting_lik}
\begin{split}
\mathcal L_e =  & \frac{1}{\sqrt{(2\pi)^{N_\text{c}} |C_d| }} \exp\left[-\frac{1}{2}(\vec d -\vec t_\text{f}-\vec e)C_d^{-1}(\vec d -\vec t_\text{f}-\vec e)\right] \\
& \times  \frac{1}{\sqrt{(2\pi)^{N_\text{c}} |C_e| }}  \exp\left[ -\frac{1}{2}\vec e C_e^{-1}\vec e \right]  \;.
\end{split}
\end{equation}
The error covariance matrix $C_e$ can be written as the direct product of the envelope $E_i$ for momentum configuration $\{i\}$ and the correlation coefficient $\rho_{ij}$
\begin{equation}
(C_e)_{ij}=E_i \rho_{ij} E_j\ \  (i,j \in [1,\ldots,N_c])\; ,
\end{equation}
where $\rho_{ii}\equiv 1$ and the off-diagonal elements account for the correlation between different configurations. The correlation coefficients need to satisfy inequalities that guarantee that the quadratic form $\vec e C_e^{-1}\vec e$ is positive semidefinite. We will employ a Gaussian correlation that is factorizable and only depends on the difference of the magnitudes of the wavenumbers
\begin{equation}
\rho_{ij}=\begin{cases}
\exp\left[-(k_i-k_j)^2 /2\Delta k^2  \right]& P\;,\\
\prod_{\alpha=1}^3 \exp\left[-(k_{i,\alpha}-k_{j,\alpha})^2 /2\Delta k^2  \right]& B \;.
\end{cases}
\end{equation}
The latter equation is unambiguous since the momenta of the bispectrum configurations are ordered $k_{i,1}\geq k_{i,2}\geq k_{i,3}$. 

Note that by fixing the correlation length our implementation of the likelihood is independent of the binning as long as the bins are sufficiently small: $k_{\rm bin}\ll \Delta k$.\footnote{Without the cross correlation coefficients, i.e. diagonal error covariance, the results would have obviously been sensitive to the binning: Unlike the statistical error which changes when one changes the size of the bins, the envelope $E(k)$ remains the same. Therefore, if all bins are uncorrelated, choosing finer and finer bins one can make the relative impact of the theoretical error smaller and smaller. The other way to see this is that finer bins allow for higher frequency functions and effectively downweight smooth error configurations expected in reality.} That is, the class of functions that are being marginalized over is fully determined by the choice of the envelope $\vec E$ and correlation length $\Delta k$ of the error.\footnote{Reference \cite{Audren:2012vy} introduces a different bin-independent method for implementing theoretical error. There, no correlation length is introduced, different components of $\vec e$ are independent, but the envelope $E(\vec k)$ is rescaled by the number of bins. This approach has two clear disadvantages: (a) By marginalizing over error functions that vary arbitrarily over different bins, a highly oscillating signal that is orthogonal to the relatively featureless gravitational uncertainties is overly penalized. (b) The significance of a signal that is coherent over $n$ correlation lengths of the error will not be enhanced by $\sqrt{n}$.} It is important to note that with sufficiently precise data one is able to constrain the theoretical error or a signal that is smaller than the envelope. For a smooth signal with characteristic scale of variation $q$ in the limit $\Delta k\ll q$ the constraints improve as the square root of the number of coherence lengths observed. In the other extreme, when $\Delta k$ is very large, the shape of the theoretical error is the same as the envelope. Marginalizing over the theoretical error is then equivalent to marginalizing over a single template $\vec E$ with a free amplitude, which can have a very small effect if the signal is orthogonal to $\vec E$. Thus, the parameter constraints in this paper \emph{do} depend on the choice of the coherence length.

In the Gaussian approximation, even with the correlated theoretical errors, it is easy to marginalize over free parameters. Integrating over $\vec e$ one can find that the final covariance matrix $C$ is simply a sum of the data covariance and the theoretical covariance
\be\label{eq:covmatrixadd}
C=\left[C_d^{-1}-C_d^{-1}\left(C_d^{-1}+C_e^{-1}\right)^{-1}C_d^{-1}\right]^{-1}=C_d+C_e\; ,
\ee
and the final likelihood is given by
\be
\label{eq:lik_with_error}
\mathcal L =\frac{1}{\sqrt{(2\pi)^{N_\text{c}} |C| }}\exp\left[-\frac{1}{2}(\vec d-\vec t) C^{-1} (\vec d-\vec t) \right]\; .
\ee
From this expression we see that the theoretical error acts as a correlated noise.

\vspace{0.5cm}
\noindent
{\em Theoretical error in data analysis: A toy example.---}So far we have discussed the theoretical error in the context of forecasts. However, everything we said is equally relevant for the analysis of the real data. For example, the theoretical error can help to avoid overfitting. To show this, let us consider a simple model in which the data are fully described by the one loop matter power spectrum ($d_k=P_\text{1loop}(k)$) and we want to measure the amplitude of matter fluctuations $A_\text{s}$ using linear theory $t_k=A_\text{s}P_\text{lin}(k)$.
Parameter constraints are derived from the maximum likelihood point
\be
\hat{A}_s =\frac{\sum_{k,k'}^{k_\text{max}} d_k\; C^{-1}_{k,k'}\;  t_{k'}}{\sum_{k,k'}^{k_\text{max}} t_k\; C^{-1}_{k,k'}\;  t_{k'}}\; ,
\ee
and the parameter error as 
\be
\Delta \hat{A}_s =\frac{1}{\sqrt{\sum_{k,k'}^{k_\text{max}} t_k\; C^{-1}_{k,k'}\; t_{k'}}}\; .
\ee
The constraints on $A_\text{s}$ as a function of the maximum wavenumber used for the fit are shown in Fig.~\ref{fig:parameter_est} with and without the theoretical error contribution to $C_{k,k'}$. Without the theoretical error, the constraint tightens quickly and soon becomes inconsistent with the true value $A_\text{s}=1$ at $k\approx0.05\ihMpc$. Once the theoretical error is taken into account, the best fit stays close to the truth. One might expect that the $\chi^2$ would have told us that the model is inconsistent with the data but in fact in our example this happens only at $k=0.09\ihMpc$. Furthermore, in a more complex setting with several free parameters the failure of the model could be masked by the freedom in parameters. We thus argue for using theoretical errors not only in Fisher matrix forecasts but also in the parameter inference algorithms.
\begin{figure}
\includegraphics[width=0.49\textwidth]{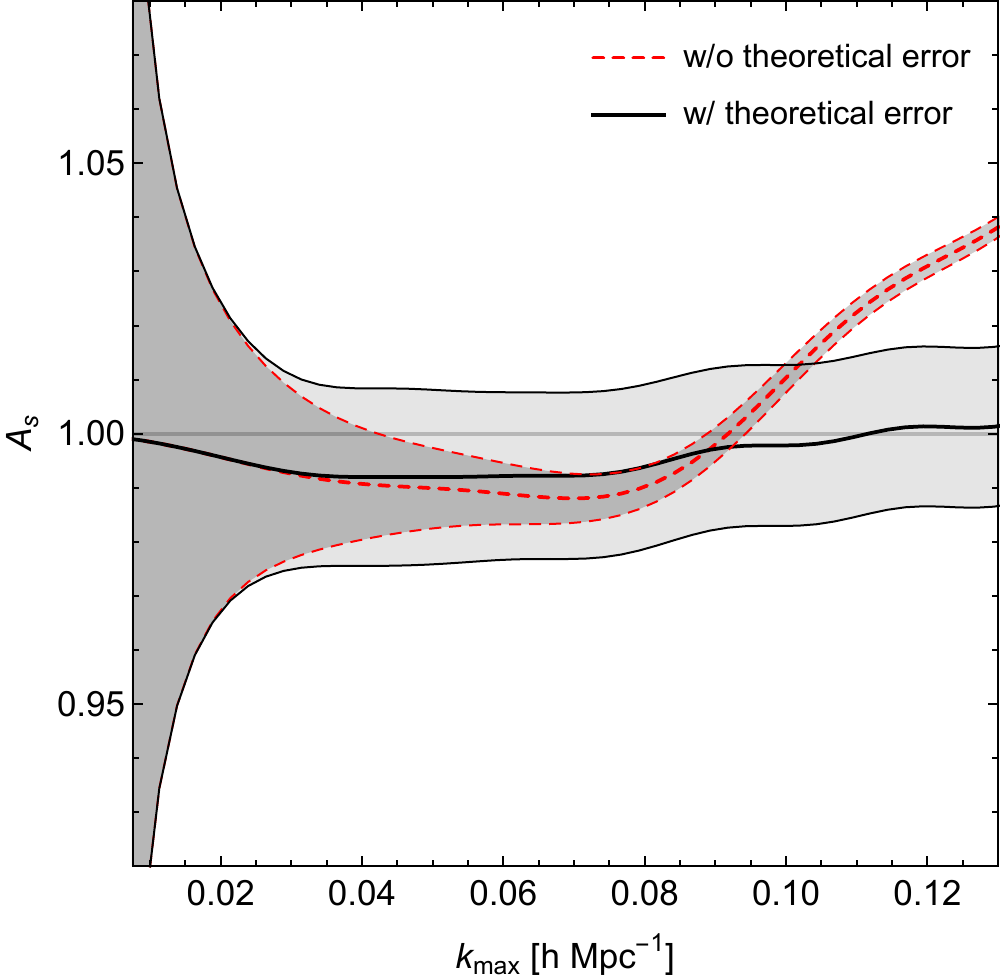}
\caption{Cumulative constraints on the power spectrum amplitude using linear theory, if the one loop matter power spectrum describes the truth. The red line with error band shows the constraint without considering the theoretical error and leads to inconsistent constraints. The black line and error bound includes the theoretical errors into the parameter estimation and leads to an unbiased estimate of the amplitude of the power spectrum. The plot is made assuming a single redshift bin at $z=0$ and an ideal survey with volume $V=(2.5\; h^{-1} {\rm Gpc})^3$.}
\label{fig:parameter_est}
\end{figure}

\subsection{The Fisher matrix}
\noindent
In this section we give explicit form of the Fisher matrix including the theoretical error both for the power spectrum and the bispectrum.

From Eq.~\eqref{eq:lik_with_error} and the definition in Eq.~\eqref{eq:fisher_definition} it follows that the power spectrum Fisher matrix is 
\be
F^p_{ij}= \sum_{z_i} \sum_{k,k'} \frac{\partial P_g(k,z_i)}{\partial p_i} (C^{-1})_{kk'} \frac{\partial P_g (k',z_i)}{\partial p_j} \;,
\ee
where the sums run over all redshift and momentum bins and $P_g$ is the theoretical galaxy power spectrum model to be described below. All terms are evaluated at the fiducial value of the parameters $\vec p_0$. Note that we will consider wide redshift bins, such that the cross spectra between bins would vanish. For surveys spanning a significant fraction of the sky one should in principle perform a decomposition of the survey into radial modes corresponding to redshift and spherical harmonics in the angular direction \cite{Heavens:1994iq}. We refrain from these complications and restrict ourselves to the flat sky analysis. For definiteness, we are using eight redshift bins with boundaries $z=0,0.3,0.6,1,1.5,2,3,4,5$. 

The galaxy power spectrum covariance matrix $C_{kk'}$ is given by
\be
C_{kk'}= \frac{(2\pi)^3}{V(z_i)}\frac{f_{\rm sky}^{-1}}{2\pi k^2 dk}\left( P_g(k,z_i)+ \frac1{ n(z_i)}\right)^2 \delta_{kk'} + (C_e)_{kk'} \;.
\ee
In this equation $V(z_i)$ is the volume of the shell that corresponds to the redshift bin $z_i$, $f_{\rm sky}$ is the observed fraction of the sky, $dk$ the width of momentum bins and $n(z_i)$ is the number density of galaxies in the redshift bin (see the appendix for a derivation of shot noise $1/n$ contribution to the galaxy power). 

Similarly, the Fisher matrix for the bispectrum is (see for example \cite{Scoccimarro:2003wn})
\be
F^b_{ij}= \sum_{z_i} \sum_{T,T'} \frac{\partial B_g(T,z_i)}{\partial p_i} (C^{-1})_{TT'} \frac{\partial B_g (T',z_i)}{\partial p_j} \;. 
\ee
The sum this time runs over all different triangles $T$ and $B_g(k_1,k_2,k_3,z_i)$ is the theoretical model for the bispectrum. All terms are evaluated at $\vec p_0$ and the ordering of the momenta is $k_1\geq k_2\geq k_3$. Therefore, the sum over triangles can be explicitly written as
\be
\sum_T \equiv \sum_{k_1=k_{\rm min}}^{k_{\rm max}}  \sum_{k_2=k_1}^{k_{\rm max}} \sum_{k_3=k_*}^{k_2} \;,
\ee
where $k_*={\rm min}(k_{\rm min},k_1-k_2)$. The covariance matrix between triangle configurations is
\be
C_{TT'}= \frac{(2\pi)^3}{V(z_i)}   \frac{\pi s_{123} f_{\rm sky}^{-1}}{dk_1 dk_2 dk_3}  \frac{M}{k_1k_2k_3}  \delta_{TT'} + (C_e)_{TT'} \;,
\ee
where $s_{123}$ is the symmetry factor that is equal to 6, 2 or 1 for equilateral, isosceles and general triangles respectively and 
\be
M = \prod_{a=1}^3\left( P_g(k_a,z_i)+ \frac1{n(z_i)}\right) \;. 
\ee
Notice that both for the power spectrum and the bispectrum we use the Gaussian approximation for the data covariance matrix $C_d$. For scales much larger than the nonlinear scale, this approximation is justified but it breaks down on small scales, where loop corrections in the input power spectrum, perturbative off-diagonal terms and eventually the one-halo term enters.

Many of the parameters $\vec p$ (for instance bias parameters) will affect both the power spectrum and the bispectrum. In order to improve the constraints on the relevant cosmological parameters one can perform a joint analysis of the power and bispectrum.  The result of this joint analysis can be assessed using the combined information from the two Fisher matrices. We perform a simple combination in which the Fisher matrix is\footnote{In principle one should include the covariance between the power spectrum and the bispectrum (as for instance calculated in \cite{Sefusatti:2006pa}). However, in the Gaussian covariance approximation employed here, this cross-correlation vanishes. It  should be included once the trispectrum contribution to the power spectrum covariance is considered. The implementation of the theoretical error for the full joint power spectrum and bispectrum analysis is then straightforward: combining the power spectrum and the bispectrum into a single data vector $\vec d=(\vec d_p, \vec d_b)$ the steps following Eq.~\eqref{eq:starting_lik} yield the full likelihood. }
\be
F= F^p + F^b + {\rm diag}(1/\sigma_{p_i}^2) \;,
\ee
where $\sigma_{p_i}$ is a prior on parameter $p_i$. In the following sections we will describe our theoretical models and specify which parameters we are using and what their priors are.

\subsection{The model}
\noindent
{\em Power spectrum.---}Let us begin with the dark matter power spectrum, including 1-loop corrections
\be
P_{\rm NL}(k,z)= A^2P(k,z) + A^4P^{\rm 1L}(k,z) + P_{\rm ct}(k,z) \;,
\label{eq:oneloopmatter}
\ee
where $A$ is a relative amplitude of the fluctuations compared to some fiducial fluctuation amplitude $\sigma_8$. The one-loop power spectrum has the usual contributions \cite{Bernardeau:2001qr}
\be
P^{\rm 1L}(k,z) = P_{22}(k,z) + P_{13}(k,z) \;.
\ee
For the one-loop calculation to be consistent we have to add a counter term with a free normalization $R_p$ \cite{Carrasco:2012cv},
\be
P_{\rm ct}(k,z) = -2A^4 R_{p}^2 \left(\frac{D_+(z)}{D_+(0)}\right)^2 k^2P(k,z)\;.
\ee 
The counter term has two pieces. One that cancels the inaccurate UV contribution from perturbation theory loop integrals. This must depend on time in the same way as the one-loop power spectrum. The second piece depends on the short scale details and in principle has an arbitrary time dependence. Here we assume the same time dependence for that part too, which is close to what is observed in simulations \cite{Baldauf:2015aha}.

The second step is to include the bias parameters. In general, the biased tracer density contrast is a functional of the dark matter fluctuations, which can be expanded in powers of fields and derivatives
\be
\delta_{g} = \sum_{\mathcal O} b_{\mathcal O} \mathcal O = b_1 \delta + \frac{b_2}{2}\delta^2 + b_{\mathcal G_2} \mathcal G_2 + \cdots \;.
\label{eq:biasexp}
\ee
The sum is over all operators $\mathcal O$ that are allowed by symmetry; they are built from the tidal tensor $\partial_i\partial_j \Phi$ \cite{McDonald:2009dh, Assassi:2014fva, Senatore:2014eva}. As an illustration, we just wrote terms up to second order in $\delta$ and leading order in derivatives. The structure of the third operator on the right hand side is $\mathcal G_2= (\partial_i\partial_j\Phi)^2-(\partial^2\Phi)^2$.

To calculate the one-loop power spectrum for biased tracers one has to consistently go to higher orders in the bias expansion. This was done in \cite{Assassi:2014fva} and in their notation
\be
\label{eq:galaxy_power_spectrum}
\begin{split}
&P_g(k,z) =  A^4 \left[ b_{\mathcal G_2}\left( b_{\mathcal G_2} - \frac 57 b_2 \right)I_{\mathcal G_2\mathcal G_2}(k,z) \right. \\
& \left. +2 b_1\left( b_{\mathcal G_2} + \frac 25 b_{\Gamma_3} \right)F_{\mathcal G_2}(k,z) + 4 b_2^2 I_{\delta_2\delta_2} (k,z)\right. \\
& \left. + 4b_1\left( b_2 - \frac 25 b_{\mathcal G_2} \right)I_{\delta_2}(k,z)\right] + b_1^2 P_{\rm NL}(k,z) + s_p(z)\;.
\end{split}
\ee
There are four different bias parameters that contribute to one-loop power spectrum: $b_1$, $b_2$, $b_{\mathcal G_2}$ and $b_{\Gamma_3}$. The explicit expressions for the functions $I_{\delta_2}$, $I_{\delta_2\delta_2}$, $I_{\mathcal G_2\mathcal G_2}$ and $F_{\mathcal G_2}$ can be found in \cite{Assassi:2014fva}. The constant $s_p$ comes from stochastic terms (see appendix). 

Finally, to $P_g(k,z)$ we have to add the effect of massive neutrinos (we denote the total mass by $M_\nu$). The main effect of neutrinos is to suppress the linear matter power spectrum. This suppression happens for wavenumbers larger than some $k_{\rm nr}$ which is the minimal comoving free-streaming wavenumber for neutrinos and depends on mass as $k_{\rm nr}(M_\nu)=0.018\sqrt{\Omega_m M_\nu/{\rm eV}} \; h{\rm Mpc}^{-1}$ \cite{Lesgourgues:2006nd}. We model this suppression in the following way
\be
\begin{split}
&P_g^{\nu}(k,z)= \kappa(k,M_\nu) A^2 b_1^2 \left( \frac{D_+(z)}{D_+(0)} \right)^{-\frac 65 f_{\nu}} P(k,z)\;,\\
&\kappa(k,M_\nu)= - 8 f_\nu \;\theta(k-k_{\rm nr}(M_\nu))\\
& \hspace{2cm} \times \left[ 1-\exp\left(-\alpha\log^2 \frac{k}{k_{\rm nr}(M_\nu)}\right) \right] \;,
\end{split}
\ee
where $f_\nu = M_\nu/(93.14\;{\rm eV}\;h^2\Omega_m)$ and $\alpha=0.12$. The function $\kappa(k,M_\mu)$ is constructed such that at very small scales $k\gg k_{\rm nr}$ it approaches $-8f_\nu$, which is a well known analytical result valid for small neutrino masses. Note that for $k > k_{\rm nr}$ the growth of perturbations is also modified and the power spectrum has slightly different time dependence \cite{Lesgourgues:2006nd}. Our model is just a rough approximation to the true shape of the neutrino contribution to the power spectrum, but it provides a good enough fit for our purposes.

\vspace{0.5cm}
\noindent
{\em Bispectrum.---}Let us now turn to the bispectrum. For the proper forecast one would have to calculate the galaxy bispectrum at one loop starting from primordial non-Gaussian initial conditions and evolve the density field keeping all relevant EFT and bias coefficients. Even for the dark matter alone, the final result at one loop is quite complicated \cite{Assassi:2015jqa}. The full analysis of the bispectrum is beyond the scope of this paper. Our primary goal is to estimate how much the theoretical uncertainties modify the constraints.

Let us again begin from the one-loop dark matter bispectrum
\be
\label{eq:bispectrum_model}
B^{\rm NL}_{123}(z)=B_{123}^{\rm tree}(z)+ B_{123}^{\rm 1L}(z) + B_{123}^{\rm ct}(z) + B_{123}^{\rm NG}(z) \;.
\ee
The first term is the tree-level bispectrum given by the following expression
\be
B^{\rm tree}_{123}(z) = 2A^4F_2({\kb}_2,{\kb}_3)P(k_2,z)P(k_3,z) + 2\; {\rm perm.}\;,
\ee  
where $F_2({\kb}_1,{\kb}_2)$ is the second order kernel of Standard Perturbation Theory (SPT).\footnote{The explicit expression for $F_2({\kb}_1,{\kb}_2)$ reads\be F_2({\kb}_1,{\kb}_2)=\frac 57 + \frac{1}{2} \mu \left( \frac{k_2}{k_1} + \frac{k_1}{k_2} \right) + \frac 27 \mu^2 \;, \ee where $\mu$ is the cosine between vectors ${\kb}_1$ and ${\kb}_2$.} The second term is the SPT 1-loop bispectrum which can be found in \cite{Scoccimarro:1997st}. The counter term for the one-loop bispectrum contains several shapes \cite{Baldauf:2014qfa,Angulo:2014tfa}, but to estimate the impact of marginalization over these additional parameters we keep just the one corresponding to the UV behavior of 1-loop SPT integral \cite{Baldauf:2014qfa}
\be
\begin{split}
B^{\rm ct}_{123}(z) = -A^6&  R_b^2\left(\frac{D_+(z)}{D_+(0)}\right)^2 k_1k_2\tilde F_2({\kb}_1,{\kb}_2)\\
& \times P(k_1,z)P(k_2,z) + 2\; {\rm perm.}
\end{split}
\ee  
In this expression $R_b$ is a free coefficient. Notice that for dark matter $R_b=R_p$. However, in the case of biased tracers this is no longer the case, because of degeneracy of these terms with derivative operators in eq. \eqref{eq:biasexp}. For the sake of generality, we keep them different from the beginning. The modified kernel $\tilde F_2({\kb}_1,{\kb}_2)$ is given by
\be
\begin{split}
(a_1+a_2\mu^2)\left( \frac{k_2}{k_1} + \frac{k_1}{k_2} \right) + \left(a_3+ \frac{k_2^2}{k_1^2} + \frac{k_1^2}{k_2^2} \right) \mu + a_4 \mu^3 \;, \nonumber
\end{split}
\ee
where $a_1,\ldots,a_4$ are numerical coefficients of order one.\footnote{The values of these coefficients can be found in \cite{Baldauf:2014qfa} \be a_1= \frac{58812}{32879},\; a_2=\frac{114624}{32879},\; a_3=\frac{231478}{32879},\; a_4=\frac{49636}{32879}.\ee}

The last term in Eq.~\eqref{eq:bispectrum_model} is due to primordial NG. The leading part of the bispectrum proportional to $f_{\rm NL}$ simply comes from the linear evolution of the initial bispectrum function and reads
\be
\label{eq:primordial_ng}
\begin{split}
B^{\rm NG}_{123}&(z) = A^4 P(k_1,z)P(k_2,z) S(k_1, k_2, k_3) \\
& \times f_{\rm NL}\cdot \frac{H_0^2\Omega_m}{D_+(z)}\frac{T(k_3)}{T(k_1)T(k_2)}k_1k_2k_3^2 + 2\;{\rm perm.} \;,
\end{split}
\ee
where the shape $S(k_1,k_2,k_3)$ is given by
\be
\begin{split}
\frac{S^{\rm eq.}_{123}}{9} = \frac1{k_1k_2^2k_3^3} - \frac1{3k_1^2k_2^2k_3^2} - \frac1{2k_1^3k_2^3} + 5\;{\rm perm.} \;,
\end{split}
\ee
for equilateral NG and 
\be
\begin{split}
\frac{S^{\rm loc.}_{123}}{3} = \frac1{k_1^3k_2^3} + \frac1{k_1^3k_3^3} + \frac1{k_2^3k_3^3}\;,
\end{split}
\ee
for local NG \cite{Creminelli:2005hu}. The amplitude $f_\text{NL}$ in two different cases is $f_{\rm NL}^{\rm eq.}$ and $f_{\rm NL}^{\rm loc.}$.

So far we have described the model for the one-loop matter bispectrum. The full calculation of one-loop bispectrum for biased tracers has not been implemented in practice and is beyond the scope of this paper. For the bispectrum, we will thus use a simple biasing model keeping only the leading terms in the bias expansion
\be
\delta_{g} = b_1 \delta + \frac{b_2}{2}\delta^2 + b_{\mathcal G_2} \mathcal G_{2}  \;.
\ee
Here we consider the first term at all orders contributing to the one loop-bispectrum (i.e. up to fourth order), whereas the second and third terms are evaluated only at tree level.
This model is incomplete (and inconsistent) at one-loop, and hence the constraints on NG that we obtain give lower bounds for the true answer. Including all relevant terms and marginalizing over the additional parameters generically weakens the constraints. In our simple biasing model we include loops only in combination with linear bias and the corresponding galaxy bispectrum is given by
\be
\begin{split}
& B_{g}(k_1, k_2, k_3, z) = b_1^3B_{123}^{\rm NL}(z) + b_1^2b_2 \Sigma_{123}(z) \\
& + 2 b_1^2b_{\mathcal G_2} \Theta_{123}(z) + s_{b,1}(z)+s_{b,2}(z)\left[P_g(k_1)+2 \text{perm}\right] \;,
\end{split}
\ee
with $\Sigma_{123}(z) = A^4P(k_1,z)P(k_2,z)+2\;{\rm perm.}$ and 
\be
\Theta_{123}(z) = A^4 \left( \frac{(\vec k_1 \cdot \vec k_2)^2}{k_1^2k_2^2} -1 \right) P(k_1,z)P(k_2,z)+2\;{\rm perm.}
\ee
The last two terms in the bispectrum come from stochastic terms (see appendix).

\vspace{0.5cm}
\noindent
{\em Theoretical error.---}The last ingredient that we need is an estimate for the theoretical error $E(k,z)$. Let us begin with the power spectrum. As we already said, we roughly expect the error to be of the form $(k/k_{\rm NL})^{(3+n)l}$. To get the correct scalings and amplitudes we fit the envelope of the explicit one-loop and two-loop calculations. The error $E(k,z)$ is given by
\be
\label{eq:power_spectrum_error}
\begin{split}
E_p(k,z)=b_1^2 \left( \frac{D_+(z)}{D_+(0)} \right)^{2l} P(k,z) \begin{cases}
(\hat k/0.31)^{1.8} & l=1\;,\\
(\hat k/0.23)^{3.3} & l=2 \;.
\end{cases}
\end{split}
\ee
where $\hat k=k/\ihMpc$. In this equation $l=1$ corresponds to the error of the linear theory and $l=2$ to the error of the one-loop power spectrum. In Fig.~\ref{fig:th_err} we show the size of these errors compared to signal for different neutrino masses. It is important to stress that our formulas are correct only for the dark matter power spectrum and that the errors for the power spectrum of biased tracers might be larger. We will use Eq.~\eqref{eq:power_spectrum_error} for all our forecasts. 

\begin{figure}
\includegraphics[width=0.48\textwidth]{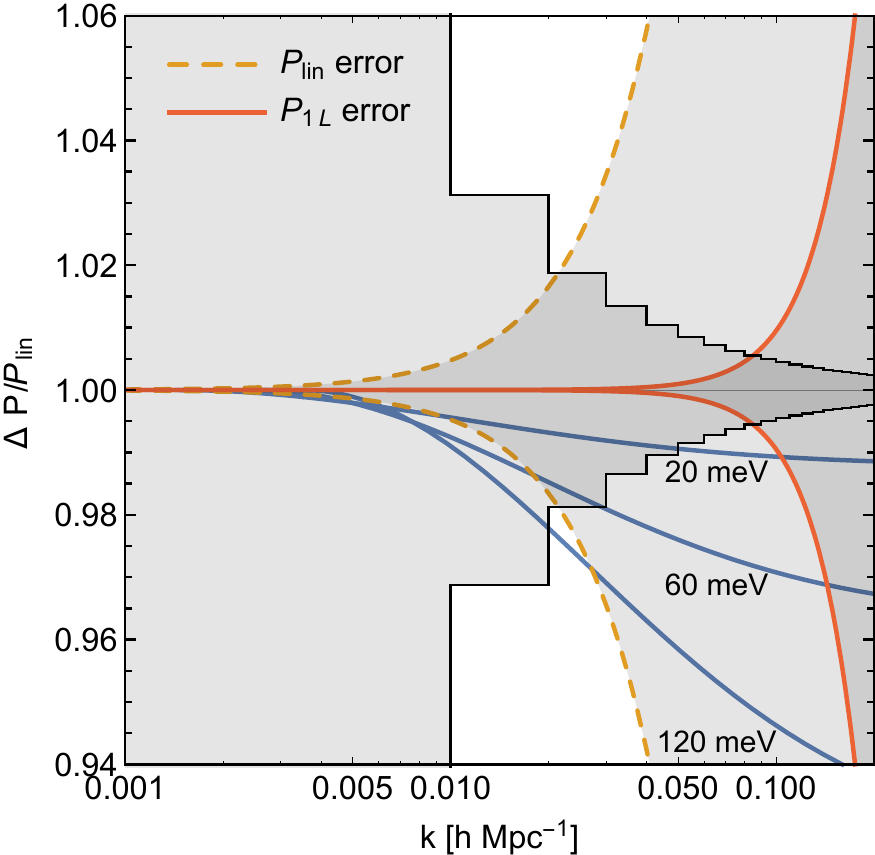}
\caption{Theoretical errors for the linear theory and one-loop power spectrum (see Eq.~\eqref{eq:power_spectrum_error}) as a function of $k$. The cosmic variance is plotted for the redshift bin $1<z<2$. Three solid lines are relative suppression of the power spectrum for three different $M_\nu$.}
\label{fig:th_err}
\end{figure}

The errors for the bispectrum are harder to estimate. We will simply assume the same power laws as in the case of the power spectrum
\be
\begin{split}
E_b & (k_1,k_2,k_3,z)=  B^{\rm tree}(k_1,k_2,k_3,z) \\
& \times 3 b_1^3 \left( \frac{D_+(z)}{D_+(0)} \right)^{2l}
\begin{cases}
(\hat k_t/3/0.31)^{1.8} & l=1\;,\\
(\hat k_t/3/0.23)^{3.3} & l=2 \;,
\end{cases}
\end{split}
\ee
where $\hat k_t=(k_1+k_2+k_3)/\ihMpc$. This is just an approximation which certainly does not capture the full shape of higher loop corrections. However, it provides a good estimate for the error. We checked it against explicit one-loop calculation of \cite{Baldauf:2014qfa} and an estimate of the two-loop bispectrum from the $N$-body simulations in the same study. As an additional check we compared our error estimate in the squeezed configuration with the approximate equations for the squeezed limit bispectrum \cite{Valageas:2013zda, Kehagias:2013paa} and found a good agreement.

\vspace{0.5cm}
\noindent
{\em Parameters and priors.---} To summarize, in our joint analysis we use the following set of parameters 
\be
\vec p = \{ f_{\rm NL}, M_{\nu}, A, R_p, R_b, b_1, b_2, b_{\mathcal G_2}, b_{\Gamma_3} \} \;.
\ee
In most of our forecasts, unless otherwise specified, we use the following fiducial values
\be
\begin{split}
\vec p_0 = \{ 0,\; & 0.06\; {\rm meV},\; 1,\; 1\; h^{-1}{\rm Mpc},1\; h^{-1}{\rm Mpc},\;\\
& \; 2,\; 0.5,\; 0.1,\; 0.1 \} \;.
\end{split}
\ee
There are no priors on $f_{\rm NL}$ and $M_{\nu}$. Priors for other parameters are 
\be
\begin{split}
& \sigma_A = 0.02\;, \quad \sigma_{b_1} = 4\;, \quad \sigma_{b_2} = 2 \;,  \\
& \sigma_{R_p} = \sigma_{R_b} =1\; h^{-1}{\rm Mpc}\;, \quad \sigma_{b_{\mathcal G_2}} = \sigma_{b_{\Gamma_3}} = 1 \;.
\end{split}
\ee
For simplicity, we assume that a single galaxy sample with specific bias parameters spans the whole range from $z=0$ to $z=5$. We are aware that this is a unrealistic scenario, but it is in line with our general approach for giving lower bounds on the errors of primordial NG. Increasing the number of free parameters can only degrade the constraints. For neutrino mass only the relatively low redshifts ($z<2$) are relevant where the results should be more robust.

We are also going to use different values of shot noise. We will always set $s_p(z)=s_{b,2}(z)=1/n(z)$ and $s_{b,1}(z)=1/n^2(z)$ with priors of $10\%$ in both cases. Here $n(z)$ is the number density of galaxies at redshift $z$. In reality, the redshift dependence should account both for the fact that distant galaxies are dimmer and that they evolve in time. Therefore, it is a function both of the survey properties, selection criteria, formation history and evolution of different types of galaxies or other tracers. To roughly get an idea how this redshift dependence affect the results, we will use a simple power law
\be
n(z) = n_0 (1+z)^{\alpha} \;,
\ee
with different values of $\alpha$. For the number density at redshift zero $n_0$, we use a range of values of $n_0= (10^{-2} - 10^{-3}) \;h^3 {\rm Mpc}^{-3} $.

In a couple of examples we will make forecast without the theoretical errors. In these cases it is important to specify what is $k_{\rm max}$ that is used. Our choice is
\be
\label{eq:kmax_definition}
k_{\rm max}(z) = 0.2\; h{\rm Mpc}^{-1} \left( \frac{D_+(z)}{D_+(0)} \right)^{-4/3} \;.
\ee
This coincides with the usual choice of $k_{\rm max}=0.2\; h{\rm Mpc}^{-1}$ at redshift zero as the scale where the perturbation theory breaks down. The time dependence is chosen to mach the evolution of the nonlinear scale for a scaling universe with $n=-1.5$.

For forecasts which include the theoretical error, $k_{\rm max}$ is automatically determined as the point at which the signal stops to grow. In order to avoid checking this condition at each step, we will always use $k_{\rm max}$ given by Eq.~\eqref{eq:kmax_definition}. We have checked that in all our examples the signal saturates below $k_{\rm max}=0.2\; h{\rm Mpc}^{-1}$.

For all our forecasts we use a sky fraction of $f_{\rm sky}=0.5$. 

\begin{figure}
\includegraphics[width=0.48\textwidth]{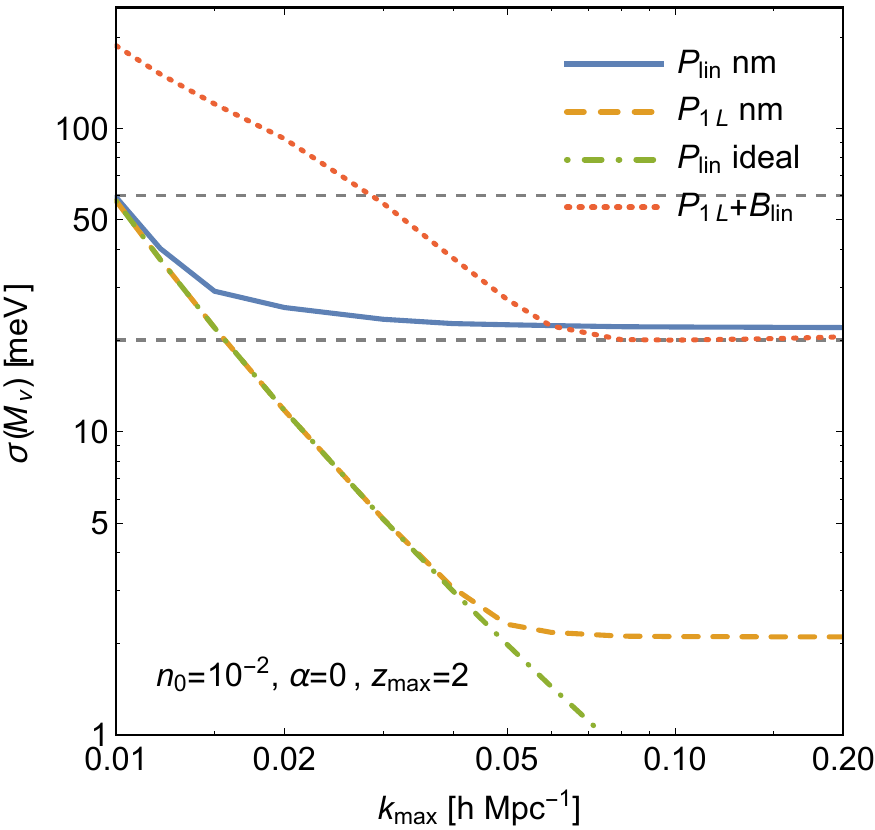}
\caption{One sigma error bar on the neutrino mass from a galaxy survey up to $z_{\rm max}=2$ as a function of $k_{\rm max}$. The two horizontal lines correspond to $M_{\nu}= 60 \;{\rm meV}$ which is the minimal mass and $M_{\nu}= 20 \;{\rm meV}$ which roughly corresponds to a $3\sigma$ detection. The solid and dashed lines are constraints without marginalization over nuisance parameters, coming from linear and one-loop power spectrum respectively with corresponding theoretical errors. The dot-dashed line is the ideal case with no theoretical errors. The dotted line is the constraint with marginalization over the EFT and bias parameters, combining the one-loop power spectrum and tree-level bispectrum and accounting for the theoretical errors. In all cases where the theoretical error is included, the constraints saturate at some $k_{\rm max}$. The constraint using the one-loop power spectrum is roughly equivalent to the ideal case with no theoretical error and shot noise $n \approx 10^{-4} \;h^3 {\rm Mpc}^{-3} $.}
\label{fig:mnkmax}
\end{figure}

\begin{figure*}
\includegraphics[width=1\textwidth]{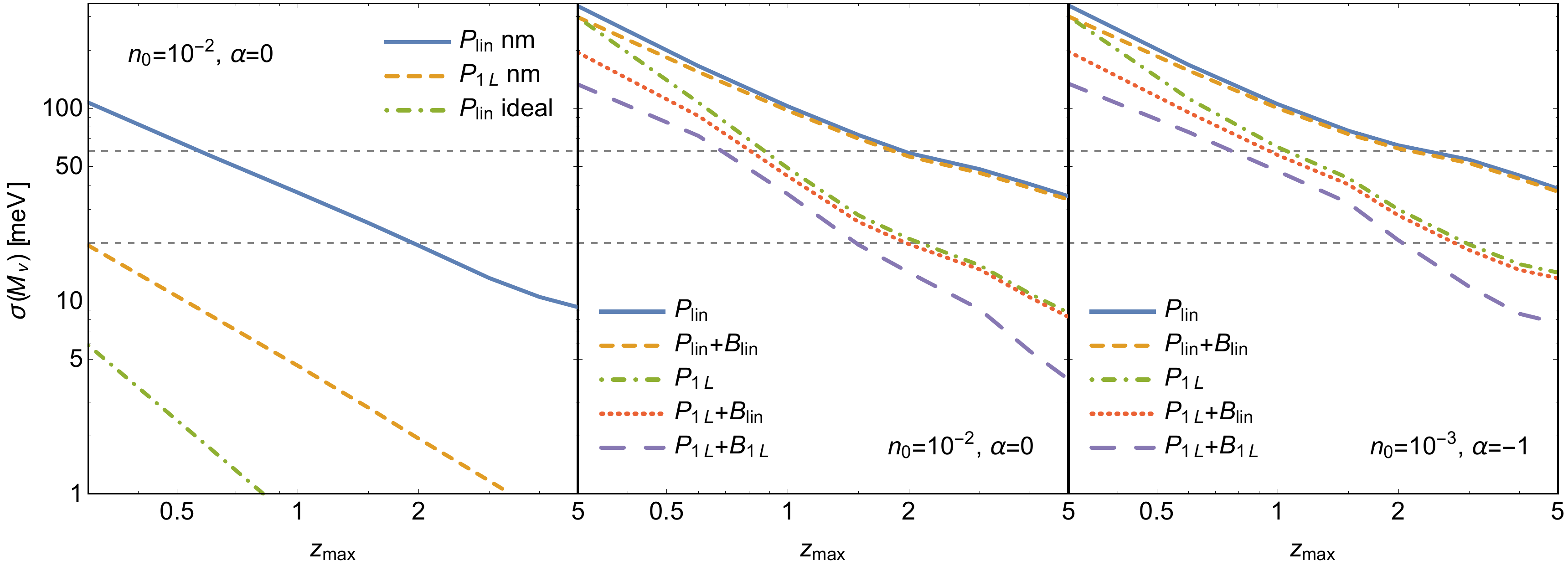}
\caption{One sigma error bar on the neutrino mass from a galaxy survey as a function of the maximal redshift $z_{\rm max}$. The two horizontal lines correspond to $M_{\nu}= 60 \;{\rm meV}$ which is the minimal mass and $M_{\nu}= 20 \;{\rm meV}$ which roughly corresponds to a $3\sigma$ detection. {\em Left panel:} Constraints without marginalization over nuisance parameters. The solid and dashed line are predictions from linear and one-loop power spectrum with corresponding theoretical errors respectively. The dot-dashed line is the ideal case with no theoretical errors. {\em Central and right panel:} Constraints with marginalization over the EFT and bias parameters for two different galaxy samples. The lines correspond to different combinations of the tree-level and the one-loop power spectrum and bispectrum accounting for the theoretical errors. The tree level bispectrum significantly improves the constraints at low redshifts and further improvements arise from the one-loop bispectrum.}
\label{fig:neutrinos}
\end{figure*}

{\section{Results}\label{sec:results}
\noindent
In this section we apply the method described above to see how much the theoretical error degrades the constraints and what are the realistic values of the sum of neutrino mass and primordial NG that one can hope to get from future surveys.

\subsection{Neutrino mass}
\noindent
We begin by constraints on neutrino mass from the galaxy power spectrum and bispectrum. The minimal neutrino mass from oscillation experiments is roughly $60\;{\rm meV}$. This minimal mass leads to a few percent level suppression of the  power spectrum around the nonlinear scale. This is of the same order of magnitude as the perturbation theory corrections. Therefore, we expect that perturbation theory can improve the constraints compared to results that one would get using the linear power spectrum only. 

In Fig.~\ref{fig:neutrinos} we show one sigma error as a function of maximal redshift $z_{\rm max}$ for a set of different perturbative schemes and assumptions. Even with no marginalization, the theoretical errors significantly degrade the constraints (by a factor of few). This is shown on the left panel. In this ideal setup, highly significant detection of the minimal neutrino mass would be possible even at low redshifts. Using the one-loop power spectrum, which has a smaller theoretical error than the linear one, makes a significant difference. Partially this is due to the increase of the range of $k$ up to which we can trust the theory, and partially due to the reduction of the error. This is illustrated in Fig.~\ref{fig:mnkmax}, where the dependence of $\sigma(M_\nu)$ as a function of $k_{\rm max}$ is shown for a survey with $z_{\rm max}=2$. Because of the theoretical error, the constraints saturate at some value of $k$, which is larger for the one-loop power spectrum. As expected, going to higher loops increases the useful number of modes. Furthermore, at the same $k$ where the constraints from the linear theory saturate, the one-loop power spectrum gives much better constraints due to the smaller theoretical error. 

Marginalization over the other parameters further degrades the constraints on neutrino mass. In particular, as it is well known, neutrino mass is highly degenerate with a combination of the amplitude of the density fluctuations $A$ and $b_1$. This degeneracy arises since the low-$k$ amplitude provides a pivot point for the suppression of power by massive neutrinos, since the shape of the transition is not sufficiently distinct. Adding the information from the bispectrum helps with breaking some of degeneracies, particularly at low redshifts. The reason is that the leading part of the bispectrum is proportional to $b_1^3A^4$, while the linear power spectrum scales as $b_1^2A^2$. In the central panel of Fig.~\ref{fig:neutrinos} we show how different combinations of input data and theoretical errors affect the constraints. The most important is the dotted red line which comes from the joint analysis of the one-loop power spectrum and the tree-level bispectrum with corresponding theoretical errors. This model is consistent in terms of bias parameters used, as described in the previous section. 

It is interesting to note that if the one-loop bispectrum is included, the constraints become even stronger. This is because our simple one-loop model for the bispectrum has the same number of bias parameters as the tree-level one. In reality the number of bias parameters would be larger and that would likely make the predictions slightly worse. Given the importance of this term, it would be interesting to do a consistent calculation of the one-loop bispectrum for biased tracers and include it in the analysis. 

At the end, the conclusion is that a significant (3$\sigma$) detection of the minimal neutrino mass seems possible even at fairly low redshifts. This is an example in which the signal is strong enough compared to the theoretical errors that going to higher orders in perturbation theory makes an important difference. For example, around $z_{\rm max}\sim 2$, including the one-loop corrections increases the significance of the detection roughly from $1\sigma$ to $3\sigma$.

Let us end with two comments. Firstly, given the Fisher matrix, it is interesting to calculate how much is $M_\nu$ degenerate with other parameters. It turns out that the degeneracies with almost all parameters are quite strong. The strongest degeneracy is with the amplitude of the power spectrum and therefore with $b_1$ and $\sigma_8$. The correlation coefficient is roughly $99\%$ at all redshifts. This indicates that the shape of the neutrino contribution to the power spectrum is not distinct enough, and that most of the information comes from the amplitude alone. Given that the theoretical error enforces relatively small $k_{\rm max}$, there is a significant degeneracy of neutrino mass with other bias parameters and $R_p$ as well. For our choice of fiducial parameters, the correlation coefficients are roughly $70-90\%$, except for $b_2$ in which case the correlation coefficient is smaller. We should also stress that different shapes that enter Eq.~\eqref{eq:galaxy_power_spectrum} for the one-loop galaxy power spectrum are highly degenerate among themselves. This is only not true for $I_{\delta_2\delta_2}$ term. The degeneracy with $s_p$ is roughly $70\%$. As an example, we give the correlation matrix with our choice of fiducial parameters with $z_{\rm max}=5$ (the coefficients have very mild redshift dependence)
\[
\begin{blockarray}{cccccccc}
M_{\nu } &  \sigma_8 &  b_1  & c_p & b_2  & b_{G_2} & b_{\Gamma _3} & R_{\text{p}}  \\
\\
\begin{block}{(cccccccc)}
1. & -0.99 & -0.99 & -0.70 & 0.21 & 0.83 & 0.84 & 0.76 \\
-0.99 & 1. & 0.999 & 0.65 & -0.15 & -0.81 & -0.83 & -0.72 \\
-0.99 & 0.999 & 1. & 0.66 & -0.18 & -0.82 & -0.83 & -0.74 \\
-0.70 & 0.65 & 0.66 & 1. & -0.67 & -0.62 & -0.59 & -0.76 \\
0.21 & -0.15 & -0.18 & -0.67 & 1. & 0.32 & 0.31 & 0.62 \\
0.83 & -0.81 & -0.82 & -0.62 & 0.32 & 1. & 0.98 & 0.93 \\
0.84 & -0.83 & -0.83 & -0.59 & 0.31 & 0.98 & 1. & 0.92 \\
0.76 & -0.72 & -0.74 & -0.76 & 0.62 & 0.93 & 0.92 & 1. \\
\end{block}
\end{blockarray}
\]

\begin{figure}
\includegraphics[width=0.48\textwidth]{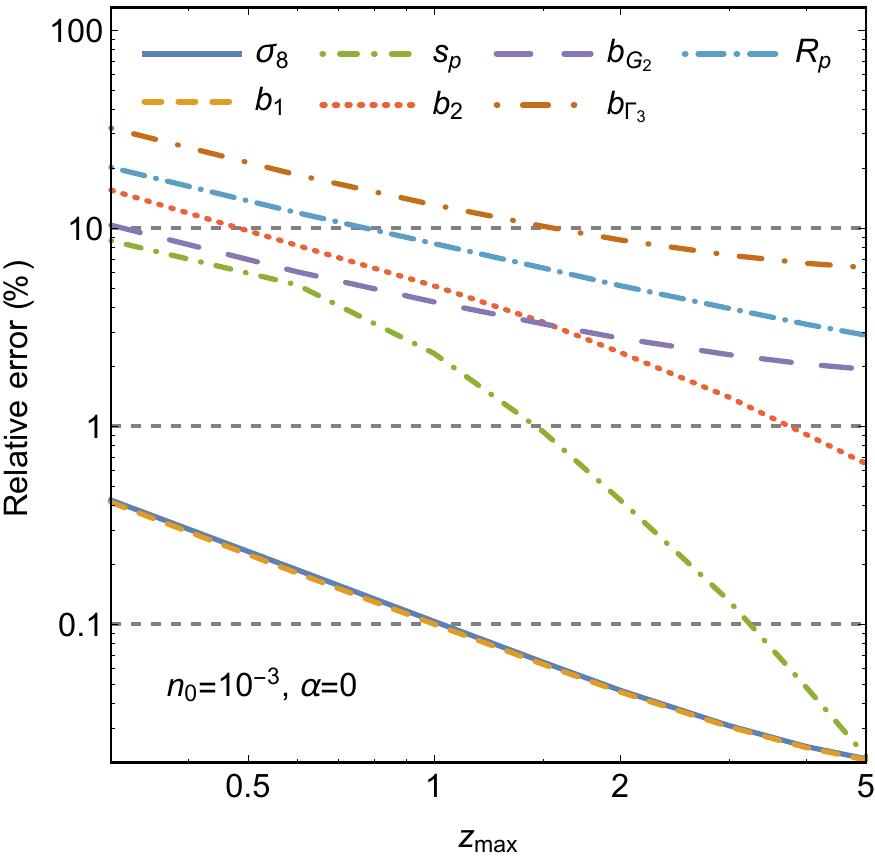}
\caption{Unmarginalized relative errors of different parameters as a function of maximal redshift $z_{\rm max}$.}
\label{fig:parameters}
\end{figure}

\begin{figure*}
\includegraphics[width=1\textwidth]{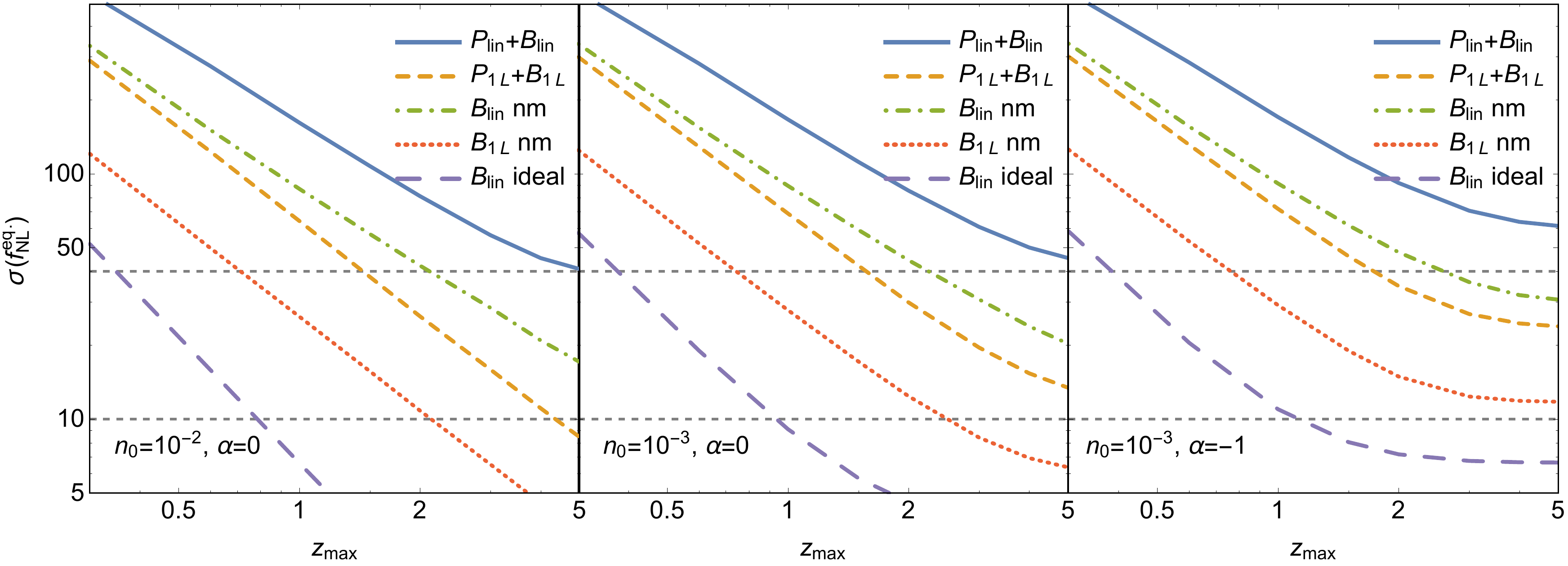}
\caption{One sigma error bar on $f_{\rm NL}^{\rm eq.}$ as a function of the maximal redshift $z_{\rm max}$. Two horizontal lines correspond to $f_{\rm NL}^{\rm eq.}= 40$ (the current strongest bound from the CMB) and $f_{\rm NL}^{\rm eq.}= 10$. Each panel shows the constraints with and without marginalization over the EFT and bias parameters. Different lines correspond to different combinations of the tree-level and the one-loop power spectrum and bispectrum. As a reference we also plot a line for the ideal case with no theoretical error and no marginalization.}
\label{fig:equilateral}
\end{figure*}

In principle, tighter priors for all parameters can be obtained from numerical simulations. The second comment is about how accurately we should know relevant bias and EFT parameters in order to get the ideal constraints on neutrino mass. One way to make an estimate is to demand that the diagonal elements of the Fisher matrix should be dominated by the priors. In other words, the priors should be smaller than the non-marginalized errors from the Fisher matrix. In Fig.~\ref{fig:parameters} we plot these non-marginalized relative errors for different parameters. For the amplitude of the power spectrum and $b_1$, which are the most important for the neutrino mass, one should have relative errors smaller than $0.1-0.5\%$ (depending on the redshift) which seems quite challenging. Other parameters, such as $b_2$, $b_{\mathcal G_2}$ or $R_p$, require precision of $1-10\%$.

{\subsection{Equilateral non-Gaussianities}\label{sec:pngresults}
\noindent
Let us now consider the constraints on primordial NG of equilateral shape. Our pNG constraints are solely obtained from the shape dependence of the tree level bispectrum and the power spectrum will be used to break degeneracies with bias parameters. We will note on explicit scale dependent bias at the end of this section.

\vspace{0.5cm}
\noindent
{\em Bispectrum.---}In Fig.~\ref{fig:equilateral} we plot $\sigma (f_{\rm NL}^{\rm eq.})$ as a function of $z_{\rm max}$ for different galaxy abundance scenarios. In the ideal case, with neither theoretical errors nor marginalization, $f_{\rm NL}^{\rm eq.}\sim 1$ can be reached at high redshift. This means that in principle there are enough modes in the perturbative regime. In practice, the theoretical error and marginalization degrade the constraints significantly. 

Including the theoretical errors only changes $\sigma (f_{\rm NL}^{\rm eq.})$ by a factor of 3 with the one-loop bispectrum and an additional factor of 3 with the tree-level bispectrum. Notice that, as in the case of neutrinos, there is a large difference between the results from the tree-level and the one-loop bispectrum. This is due to the fact that including higher loops increases $k_{\rm max}$ and reduces the error for $k<k_{\rm max}$.

Marginalization degrades the constraints by additional factor of few. This is not surprising given that the gravitational contributions are not very orthogonal to the equilateral shape. With our simple model for the one-loop bispectrum of biased tracers, the current Planck limits can be reached with a survey that would map the distribution of galaxies up to redshift $z\sim 1.5$. With a more realistic model which will contain more bias parameters, the results are expected to get weaker. Going to higher redshifts, our analysis indicates that reaching $f_{\rm NL}^{\rm eq.} \sim 10$ will be very challenging. 

\begin{figure*}
\includegraphics[width=1\textwidth]{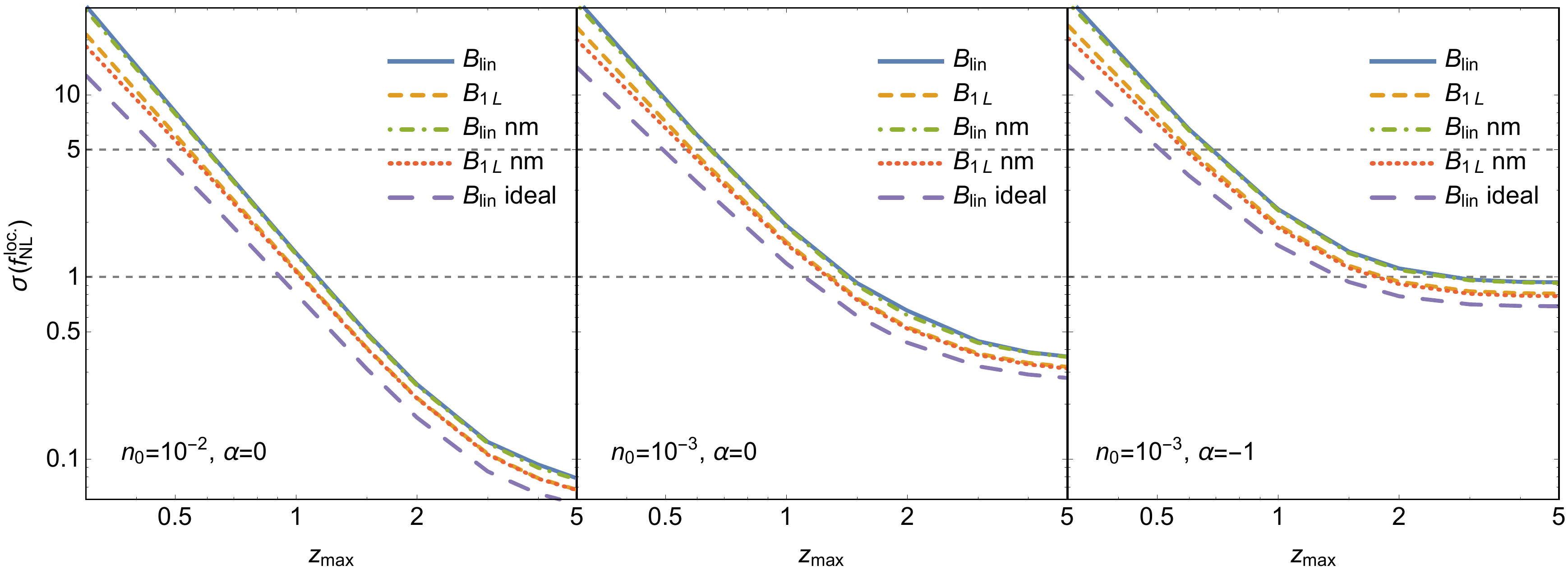}
\caption{One sigma error bar on $f_{\rm NL}^{\rm loc.}$ as a function of the maximal redshift $z_{\rm max}$. Two horizontal lines correspond to $f_{\rm NL}^{\rm loc.}= 5$ (the current strongest bound from the CMB) and $f_{\rm NL}^{\rm loc.}= 1$ which is an interesting theoretical threshold. Each panel shows the constraints with and without marginalization over the EFT and bias parameters. Different lines correspond to different combinations of the tree-level and the one-loop bispectrum and corresponding errors. The effects of the marginalization are minimal, given that the local shape is orthogonal to gravitational contributions. We also plot as a reference a line for the ideal case of no theoretical error and no marginalization.}
\label{fig:local}
\end{figure*}

\vspace{0.5cm}
\noindent
{\em Scale dependent bias.---} Equilateral NG do not affect only the bispectrum. They can also contribute to the power spectrum through a scale dependent bias of the form 
\be
\Delta b_1(k) \approx 9(b_1-1)  f_\text{NL}^\text{eq.} \cdot \Omega_m \delta_c \frac{ H_0^2R^2(z)}{D_+(z)T(k)} \;.
\ee
(This form can be obtained by taking the squeezed limit $k_1\ll k_{2,3}$ of \eqref{eq:primordial_ng} as a correction to the power of short scale modes $k_{2,3}$ with the characteristic size $R(z)$, the Lagrangian size of objects observed at redshift $z$. $b_1-1$ and $\delta_c = 1.686$ typically appear in the simplest halo models that relate the change in the power to the bias parameters \cite{Schmidt:2010gw}.) We choose the same time dependence as for the counter terms in the power spectrum: $R(z)=R_0 D_+(z)/D_+(0)$. The power spectrum is modified in the following way
\be
\label{eq:scale_dep_eq_pw}
P_g(k, z) = (b_1+\Delta b_1(k))^2 P(k,z) \;,
\ee
and one can put constraints on $f_\text{NL}^\text{eq.}$ measuring its shape carefully. However, the amplitude of $\Delta b_1(k)$ is very small, typically $R^2 H^2\sim 10^{-6}$. Note that compared to the similar term in the bispectrum, the effect of the scale dependent bias at some scale $k$ is $R^2k^2$ times smaller. For perturbative scales $R k < 1$, and we expect weaker limits on $f_\text{NL}^\text{eq.}$ than what we get from the three-point function. 

To test this expectation we do a simple forecast using just the model described by Eq.~\eqref{eq:scale_dep_eq_pw}. We do not include the theoretical error and we do not marginalize over $b_1$. For example, the choice of $R_0=3\; h^{-1}{\rm Mpc}$ and the same $k_{\rm max}$ as before leads to $\sigma(f_\text{NL}^\text{eq.}) = 12$ at redshift $z_{\rm max}=1.5$. This should be compared to the ideal case from the bispectrum analysis at Fig.~\ref{fig:equilateral}. Obviously, the bispectrum constraints are stronger. 

The result strongly depends on the choice of $R_0$. The constraint on equilateral NG naively scales as $\sigma(f_\text{NL}^\text{eq.}) \sim R_0^{-2}$. Choosing a larger $R_0$ (which corresponds to larger haloes) seems to reduce the error significantly. However, at the same time, the value $k_{\rm max}$ has to be smaller. In order to stay in the perturbative regime, we cannot use the modes with wavelengths shorter than the size of the halo. Therefore $k_{\rm max}<R_0^{-1}$, and higher $R_0$ leads to smaller number of modes.

The constraints are further degraded by marginalizing over other parameters and including the theoretical error. The scale dependence of $\Delta b_1(k)$ is not protected by symmetries and it is degenerate with loop and higher derivative corrections. Indeed, for large $k$ the transfer function scales as $T(k)\sim k^{-2} \log k$. For example, even a simple extension of the model including the one-loop contributions proportional to $k^2$
\be
P_g(k, z) = (b_1+\Delta b_1(k))^2 P(k,z) (1+ R_p^2 k^2) \;,
\ee
degrades the constraints on $f_\text{NL}^\text{eq.}$ significantly, after marginalization over $b_1$ and $R_p$. For example, at redshift $z_{\rm max}=1.5$, the constraints are $\sigma(f_{\rm NL}^{\rm eq.})\approx800$ and $\sigma(f_{\rm NL}^{\rm eq.})\approx450$ for the linear and the one-loop power spectrum respectively. The full model for the power spectrum, once other parameters are included, leads to even worse constraints. In order to get results competitive with the bispectrum analysis, one would have to use $R_0\approx 10 \; h^{-1}{\rm Mpc}$ with the same $k_{\rm max}$. 

Using the scale dependent bias and perfect knowledge of the power spectrum up to $k=0.2\;\ihMpc$, \cite{Dore:2014cca} forecasted constraints of $\sigma(f_\text{NL}^\text{eq.})\sim 7$ for $z_{\rm max}=1.5$ and marginalizing over bias parameters. For reasons we explained here, we believe that this number is optimistic. Given the importance of the question, this analysis requires further investigation.

\begin{figure*}
\includegraphics[width=1\textwidth]{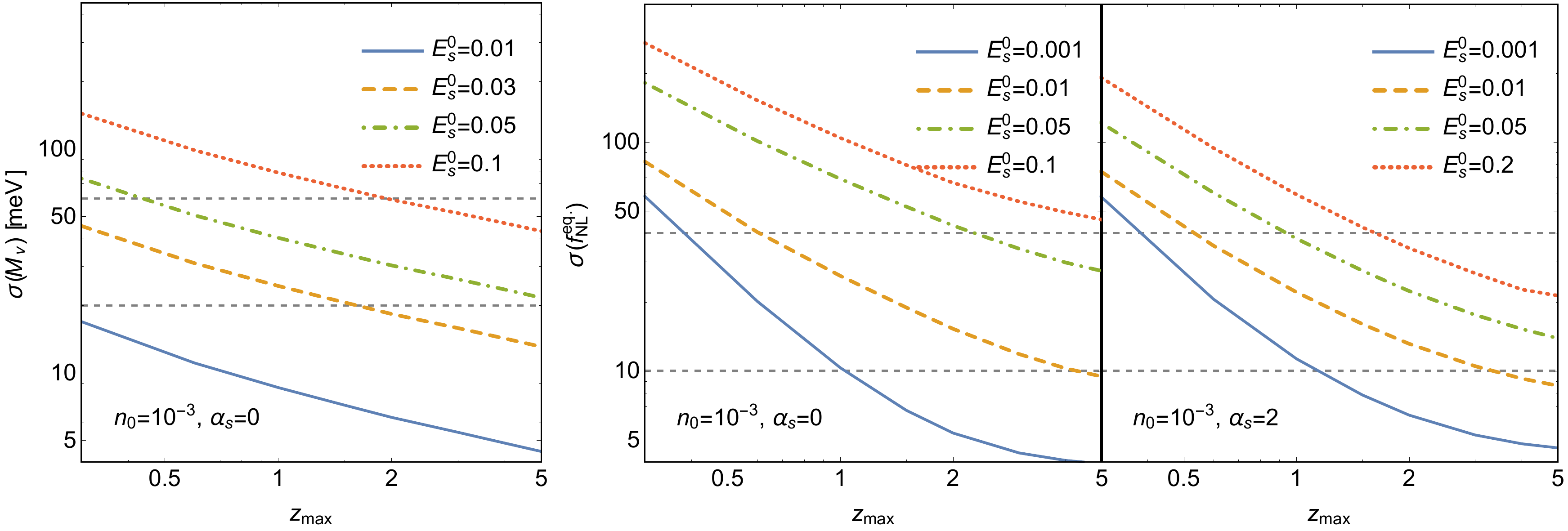}
\caption{{\em Left panel:} One sigma error bar on the neutrino mass as a function of the maximal redshift $z_{\rm max}$ for different values of $E_s^0$, keeping fixed the number density of galaxies $n_0 = 10^{-3}\;h^3 {\rm Mpc}^{-3}$ at all redshifts and $\alpha_s=0$. A $3\sigma$ detection of the minimal neutrino mass can be achieved with $E_s^0\sim {\rm few}\%$. {\em Central and right panel:} One sigma error bar on $f_{\rm NL}^{\rm eq.}$ as a function of the maximal redshift $z_{\rm max}$ for different values of $E_s^0$ and $\alpha_s$, keeping fixed the number density of galaxies $n_0 = 10^{-3}\;h^3 {\rm Mpc}^{-3}$ at all redshifts. Two horizontal lines correspond to $f_{\rm NL}^{\rm eq.}= 40$ (the current strongest bound from the CMB) and $f_{\rm NL}^{\rm eq.}= 10$. In order to significantly improve current upper bounds one would need simulations with $0.1\%$ precision at $k=0.2\; h{\rm Mpc}^{-1}$, irrespectively of the value for $\alpha_s$.}
\label{fig:simulations}
\end{figure*}

\subsection{Local NG}
\noindent
The issues with the theoretical error we discussed so far in principle apply to local NG too. However, the prospects of constraining local NG from the LSS are much brighter. This is possible thanks to a number of nonperturbative results, based on the equivalence principle, which allow us to use information even from the nonlinear regime of LSS. We briefly describe two ways to measure $f_{\rm NL}^{\rm loc.}$ and check whether from the bispectrum alone one can reach the theoretically interesting target of $f_{\rm NL}^{\rm loc.}\sim1$.

\vspace{0.5cm}
\noindent
{\em Bispectrum.---} In the presence of the local NG the squeezed limit bispectrum scales as
\be
B(q, k, k')|_{q\to0} \sim P(q)P(k) \cdot \frac{3 f_{\rm NL}^{\rm loc.} \Omega_m}{D_+(0)} \frac{H_0^2}{q^2} \;.
\ee
This is a result of perturbation theory, but similarly to the scale dependent bias, this shape of the squeezed limit of the bispectrum is protected by the equivalence principle. Including biased tracers or going beyond the nonlinear scale for the short modes cannot generate the characteristic $1/q^2$ scaling. 

In the context of perturbation theory and its theoretical errors the above scaling implies that even if we have  poor theoretical control at short distances, sufficiently squeezed triangles still contribute to the signal-to-noise. This is very different from equilateral NG. Our goal here is to estimate how well one can constrain $f_{\rm NL}^{\rm loc.}$ using the information from the bispectrum only (without the scale dependent bias) and with modes in the perturbative regime. 

As expected, our results show that the marginalization does not do almost any damage because the local shape is quite orthogonal to gravitational contributions. Including the theoretical error degrades constraints only slightly. For these reasons, the results are not very different from the ideal case and $f_{\rm NL}^{\rm loc.}\sim 1$ seems to be within the reach of futuristic galaxy surveys. In a more realistic forecast, which includes the scale dependence of bias parameters and possible extension to even shorter scales (with appropriate covariance matrix), the signal can only increase and that would further improve the final bounds. Let us note that our forecast is not very reliable for $f_{\rm NL}^{\rm loc.}< 1$, where the relativistic corrections must be also taken into account. However, this can be done straightforwardly and we do not expect it to change the result significantly.

\vspace{0.5cm}
\noindent
{\em Scale dependent bias.---} Similarly to equilateral NG, the correction to the linear bias coefficient due to the local NG is scale dependent with a particular behavior $\Delta b_1\sim f_{\rm NL}^{\rm loc.} H_0^2/q^2$ \cite{Dalal:2007cu}, in the limit $q$ goes to zero. This kind of momentum dependence cannot be generated by any astrophysical processes. In single-field models of inflation this scale-dependent bias vanishes \cite{dePutter:2015vga,Dai:2015jaa}, and therefore it is a powerful probe for distinguishing different inflationary models. This result is exact and the constraints are dominated just by statistical uncertainties. Therefore, the usual forecasts are reliable and $f_{\rm NL}^{\rm loc.}\sim1$ is achievable even at relatively low redshifts (see \cite{Dore:2014cca}).

\subsection{Simulations}
\noindent
{\em Simulations.---} Finally, we comment on the kind of precision needed in numerical simulations to make a significant detection of the minimal neutrino mass or a large improvement on the upper bounds for equilateral NG. The precision of the power spectrum from dark matter only simulations is currently $\mathcal O(1\%)$ \cite{Heitmann:2008eq,Schneider:2015yka}. The precision can be probably improved, at least on large scales, using hybrid schemes which combine the perturbation theory and N-body simulations (see for example \cite{Tassev:2015mia}). However, this is not sufficient because on top of the dark matter distribution one has to add a biasing model that introduces additional errors, in the same way  as for perturbation theory.

Alternatively, one can imagine that the numerical simulations will improve so much in the future that they will be able to simulate galaxy formation and therefore directly provide the power spectrum or the bispectrum for galaxies. In this idealistic setup, with essentially no free parameters, the only degradation of the constraints comes from the simulation error. We parametrize the power spectrum error in the following way
\be
E_s (k) = E_s^0\left(\frac{k}{0.2 \; h{\rm Mpc}^{-1}} \right)^{\alpha_s} P(k,z) \;.
\ee
We will use the same relative error for the bispectrum. We choose to normalize momenta to $k=0.2 \; h{\rm Mpc}^{-1}$ which corresponds to $k_{\rm max}$ at redshift zero which we use in our forecasts. The simulations can certainly be used even at higher $k$, but then one has to go beyond the Gaussian covariance matrix. This would add some additional information, until the modes become highly correlated. In this sense our estimates are slightly pessimistic. 

In Fig.~\ref{fig:simulations} we show $1\sigma$ errors on neutrino mass and equilateral NG as a function of the maximal redshift $z_{\rm max}$ for different choices of $E_s^0$ and $\alpha_s$. Even with a constant relative error ($\alpha_s=0$) it is sufficient to have a precision $E_s^0\sim\mathcal O(1\%)$ in order to significantly detect the minimal neutrino mass. The situation with equilateral NG is quite different. For the constant relative error, one needs at least $E_s^0\sim \mathcal O(10^{-3})$ to obtain a significant improvement compared to the CMB limits. Changing $\alpha_s$ does not change this result. In conclusion, reaching $f_{\rm NL}^{\rm eq.}\sim 1$ is very challenging even with simulations.

{\section{CMB Lensing}
\noindent 
Another way to constrain the sum of neutrino masses is through the weak gravitational lensing of the CMB. Disadvantages of lensing are that the number of available modes is much smaller than in a 3D survey and that one can measure only the integrated mass along the line of sight. Nevertheless, CMB lensing is perhaps a cleaner probe of the matter power spectrum than galaxy clustering or galaxy weak lensing since it for example it does not suffer from intrinsic alignments. In this section we repeat forecast for constraints on neutrino mass from CMB lensing, including theoretical errors.

\subsection{The model}
\noindent
The deflection potential in the Limber approximation can be expressed as the integral along the line of sight \cite{Lewis:2006fu}
\be
\label{eq:deflection_pot}
C_l^{dd}=4 \int \frac{d \chi}{\chi^2}\; \left(\frac{\chi -\chi_\text{s}}{\chi \chi_\text{s}}\right)^2 P_\phi(l/\chi,\chi) \;,
\ee
where $\chi_\text{s}$ is the geodesic distance to the CMB and $P_{\phi}(\vec k, \eta)$ the power spectrum of the gravitational potential.\footnote{Note that our definition of the power spectrum is $\langle \phi(\vec k) \phi(\vec k') \rangle= (2\pi)^3\delta(\vec k -\vec k')P_\phi(k)$.} The relation of $P_{\phi}(\vec k, \eta)$ to the matter power spectrum is
\be
P_\phi(k,z) = \frac 9 4 \frac{\Omega_m H_0^4}{k^4 a^2(z)} P_{\rm NL}(k,z) \;.
\ee
Therefore, given the model for $P_{\rm NL}(k, z)$, it is straightforward to calculate the model for the deflection potential. For a calculation of the one- and two-loop contributions to the lensing power spectrum see \cite{Foreman:2015uva}. 

Our model is based on the one-loop matter power spectrum in Eq.~\eqref{eq:oneloopmatter}. However, this model has to be completed. The reason is that the validity of the one-loop expression for the matter power spectrum is restricted to wavenumbers well below the non-linear scale $k_{\rm NL}$ while the integral in Eq.~\eqref{eq:deflection_pot} picks up a small contribution from non-linear scales even for comparably small $l$. To obtain a realistic shape for the deflection power spectrum we have to account for the non-linear power on scales smaller than $k_{\rm NL}^{-1}$. This non-linear power can not be accounted for by the theoretical error discussed so far, since by construction the envelope of the higher order perturbative loops is restricted to $k < k_\text{NL}$ as well. Before we discuss the concrete model, let us briefly discuss the phenomenology of the non-linear power spectrum: close to $k_\text{NL}$ the perturbative part goes to zero and the stochastic part of the power spectrum kicks in. The latter part can not be circumvented, no matter how many loops are considered. We work with the following model
\be
\begin{split}
&P_{\rm NL}(k, z) = P(k,z)+(P_{\rm 1L}+P_{\rm ct})W(k,k_{\rm NL}) \\
&\quad + P_{\rm fit}W(k,k_1)(1-W(k,k_{\rm NL})) + P_{\rm stoch.} \;,
\end{split}
\ee 
where $W(k,k')=\exp[-(k/k')^2(D_+(z)/D_+(0))^{8/3}]$, $k_{\rm NL}=0.3\;h{\rm Mpc}^{-1}$ and $k_1=1\;h{\rm Mpc}^{-1}$. The first line is the perturbation theory expression up to one-loop. The stochastic part of the power spectrum, which is uncorrelated to the perturbation theory result, can be written like
\be
P_{\rm stoch.} = 480\pi \left( \frac{D_+(z)}{D_+(0)} \right)^{5.5}  \left( \frac{k}{k_s} \right)^{4} \frac1{(1+(k/k_s)^2)^3} \;,
\ee
where $k_s= (D_+(z)/D_+(0))^{-1.5}\cdot 0.8\;h{\rm Mpc}^{-1}$. Finally, we have to add a term that interpolates between the one-loop and the stochastic term. This term contains all higher loop contributions but for simplicity we use an expression which fits the nonlinear power spectrum well in the range of scales and redshifts that dominantly contribute to the lensing signal
\be
P_{\rm fit} (k,z) = 320 k^{-0.5} \left( \frac{D_+(z)}{D_+(0)}\right)^4 \;.
\ee

This model is just a simple fit that we use in order to roughly calculate the lensing power spectrum. We have verified this model with outputs of the CAMB Halofit implementation of the non-linear power spectrum \cite{Lewis:1999bs}.
The dominant contribution to the final result comes from the linear theory for every $l$ of interest. For example, the corrections to the linear prediction are roughly $15\%$ at $l=1000$. The details of the matching to the perturbation theory and the stochastic term are not very relevant for our results and we include the uncertainties in these terms in our theoretical error.  

The theoretical error on $C_l^{dd}$ can be estimated by using Eq.~\eqref{eq:deflection_pot}. Like the amplitude, the error also has two contributions. In the perturbative regime we use the usual equation \eqref{eq:power_spectrum_error} for the error of the tree-level and the one-loop power spectra. Beyond $k_{\rm NL}$ the errors blow up, so in order not to overestimate their contribution we use the perturbative estimates only at wavenumbers for which the relative error is smaller than a certain threshold $\epsilon$. Beyond that we use constant relative error equal to $\epsilon$. We will consider two different choices $\epsilon= 5\%$ and $\epsilon= 10\%$. In this way, we allow for fairly large uncertainty in the matter power spectrum for $k>k_{\rm NL}$. In Fig.~\ref{fig:lensing} we show the impact of these uncertainties on the deflection power spectrum. 

\begin{figure}
\includegraphics[width=0.48\textwidth]{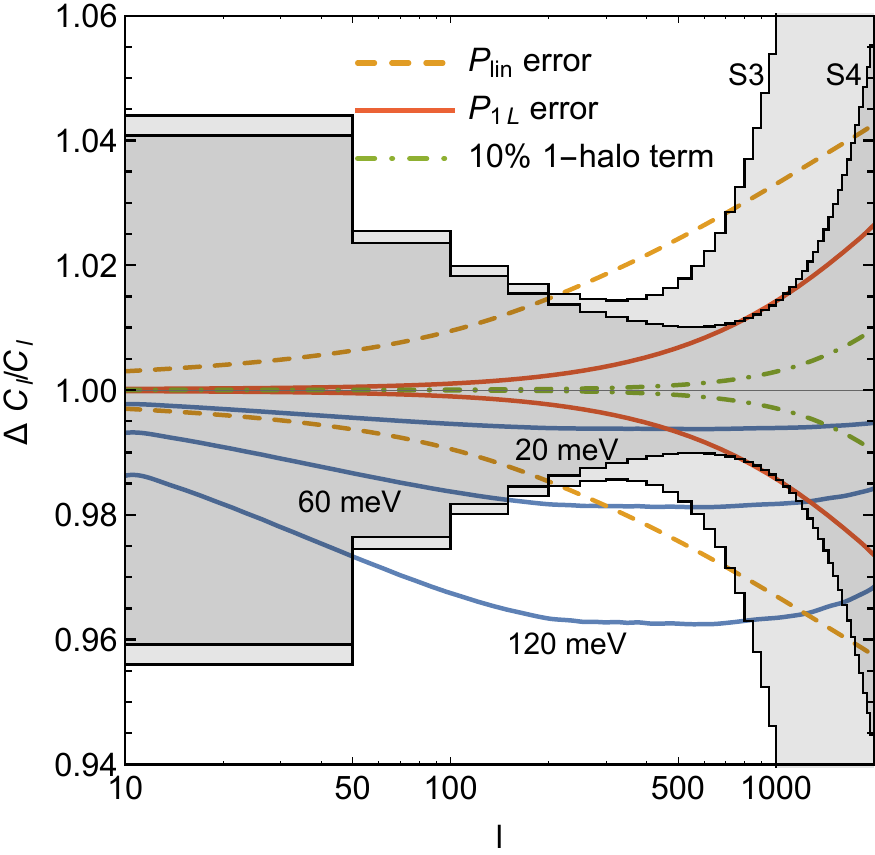}
\caption{Relative error and amplitude of various terms compared to the deflection power spectrum. The 1-loop and 2-loop errors are calculated using $\epsilon=0.05$. }
\label{fig:lensing}
\end{figure}

Finally, in our forecasts we have to include the statistical errors in the measurement of $C_l^{dd}$. The deflection angles are not directly observable but have to be reconstructed from the temperature and polarization maps. The total error is the sum of the cosmic variance and the instrumental noise
\be
\Delta C_l^{dd} = \frac 1{\sqrt{l\Delta l f_{\rm sky}}} (C_l^{dd} + N_l^{dd}) \;,
\ee
where $\Delta l$ is the bin width. The instrumental noise for each of these is given by \cite{Knox:1995dq}
\be
\begin{split}
N_l^{TT} & =\frac{\Delta_T^2}{T_\text{CMB}^2} \exp\left[\frac{l(l+1)\theta^2}{8\log 2}\right] \;, \\
N_l^{EE} & = N_l^{BB} =\frac{2\Delta_T^2}{T_\text{CMB}^2} \exp\left[\frac{l(l+1)\theta^2}{8\log 2}\right] \;,
\end{split}
\ee
where $\Delta_T=\sigma \theta$, $\sigma$ is the pixel noise variance and $\theta$ the FWHM beam size. The minimum variance estimator for the deflection potential and its error were calculated in \cite{Hu:2001kj}. We use their formalism to find $N_l^{dd}$, which comes from an optimal combination of all polarization and temperature measurements. For currently ongoing and future polarization experiments with high sensitivity, the dominant (minimal variance) contribution comes from the correlations between primordial E-modes and the B-modes generated by gravitational lensing of the primordial E-modes. 

In Fig.~\ref{fig:lensing} we show the relative contribution of different terms to the power spectrum. The noise is calculated for two different classes of experiments. For the stage III (S3) type experiment we use the following parameters
\be
f_{\rm sky} = 0.5\;, \quad \theta = 1\;{\rm arcmin}\;, \quad \Delta_T = 8\; \mu{\rm K\;arcmin}\;,
\ee
while the stage IV (S4) type experiment is characterized by (see for example \cite{Allison:2015qca})
\be
f_{\rm sky} = 0.5\;, \quad \theta = 3\;{\rm arcmin}\;, \quad \Delta_T = 1\; \mu{\rm K\;arcmin}\;.
\ee
On large scales (low-$l$) the cosmic variance dominates, whereas on small scales (high-$l$) the instrumental noise dominates. This leaves only a fairly small window, where the percent level effects exceed the observational error bars.
The minimal neutrino mass $M_\nu = 60 \;{\rm meV}$ has a $2\%$ effect and is strongly degenerate with the amplitude of the power spectrum over the range where its signal exceeds the error bars. The one-loop theoretical error has a similar size for the relevant scales, while the two-loop error is significantly smaller. 

\begin{table}
\begin{center}
\begin{tabular}{c c c c c}
\hline
{$\sigma_{A}[\%]$} & {\bf Lin$_{\epsilon=0.05}$} & {\bf 1L$_{\epsilon=0.05}$} & {\bf Lin$_{\epsilon=0.1}$} & {\bf 1L$_{\epsilon=0.1}$} \\
\hline
$1.0\;$ & $73\; {\rm meV}\;\;$  & $64\; {\rm meV}\;\;$ & $80\; {\rm meV}\;\;$ & $65\; {\rm meV}\;\;$ \\
$0.5\;$ & $48\; {\rm meV}\;\;$  & $37\; {\rm meV}\;\;$ & $57\; {\rm meV}\;\;$ & $38\; {\rm meV}\;\;$ \\
$0.1\;$ & $35\; {\rm meV}\;\;$  & $20\; {\rm meV}\;\;$ & $46\; {\rm meV}\;\;$  & $22\; {\rm meV}\;\;$ \\
\hline
\end{tabular}
\caption{$1\sigma$ errors for the fiducial sum of neutrino masses $M_{\nu}=60 \; {\rm meV}$ for a S3 like experiment.}
\label{tab:neutrinos-III}
\end{center}
\end{table}

\begin{table}
\begin{center}
\begin{tabular}{c c c c c}
\hline
{$\sigma_{A}[\%]$} & {\bf Lin$_{\epsilon=0.05}$} & {\bf 1L$_{\epsilon=0.05}$} & {\bf Lin$_{\epsilon=0.1}$} & {\bf 1L$_{\epsilon=0.1}$} \\
\hline
$1.0\;$ & $72\; {\rm meV}\;\;$  & $62\; {\rm meV}\;\;$ & $79\; {\rm meV}\;\;$ & $64\; {\rm meV}\;\;$ \\
$0.5\;$ & $45\; {\rm meV}\;\;$  & $35\; {\rm meV}\;\;$ & $55\; {\rm meV}\;\;$ & $37\; {\rm meV}\;\;$ \\
$0.1\;$ & $32\; {\rm meV}\;\;$  & $18\; {\rm meV}\;\;$ & $43\; {\rm meV}\;\;$  & $20\; {\rm meV}\;\;$ \\
\hline
\end{tabular}
\caption{$1\sigma$ errors for the fiducial sum of neutrino masses $M_{\nu}=60 \; {\rm meV}$ for a S4 like experiment.}
\label{tab:neutrinos-IV}
\end{center}
\end{table}

It is important to stress that the contribution of the stochastic term to the theoretical error is small compared to the instrumental noise. This is important for two reasons. Firstly this means that using the perturbation theory it is possible to further reduce the error. Given the difference of the one-loop and two-loop envelopes, it is reasonable to expect that the error for the two-loop power spectrum is significantly smaller than the noise of both S3 and S4 experiments. In that regime the theoretical error only slightly changes the usual forecasts (see for instance \cite{Allison:2015qca}). The second reason is that the size of the stochastic contribution can be used to estimate the effects of baryons on the lensing potential. The contribution of baryons on large scales can be captured in the EFT framework \cite{Lewandowski:2014rca}, but on small scales it is beyond the reach of perturbation theory. Given the smallness of the dark matter stochastic term, we do not expect the baryons to contribute significantly to the theoretical error.

\subsection{Results}
\noindent 
Once the model, the theoretical error and the noise for the power spectrum are known, it is straightforward to do the forecast including the theoretical uncertainties. The set of parameters we use is
\be
\vec p = \{M_{\nu}, A, R_p \} \;,
\ee
with the following fiducial values
\be
\begin{split}
\vec p_0 = \{ 60\; {\rm meV},\; 1,\; 1\; h^{-1}{\rm Mpc} \} \;.
\end{split}
\ee

In Tab.~\ref{tab:neutrinos-III} and Tab.~\ref{tab:neutrinos-IV} we summarize our results. We give the $1\sigma$ errors on the minimal neutrino mass $M_{\nu}=0.06\;{\rm eV}$. The main degeneracy of $M_\nu$ is with the amplitude of the power spectrum. Therefore, the results strongly depend on the prior on $A$. We use there different values $\sigma_A = 0.01$, $\sigma_A = 0.005$ and $\sigma_A = 0.001$. The prior for the EFT parameter is $\sigma_{R_p} = 0.1 \; h^{-1}{\rm Mpc}$.

Different columns are results for the linear and the one-loop power spectrum for two different choices of $\epsilon$. As expected from Fig.~\ref{fig:lensing}, using the one-loop power spectrum improves the constraints, and with sufficiently tight prior on $A$ a significant detection of the minimal neutrino mass is possible. 

The constraints are almost the same for S3 and S4 type of experiments. This is somewhat surprising given that the noise for S4-like experiment is significantly lower (see Fig.~\ref{fig:lensing}). For the linear theory the reason is that the theoretical error dominates the noise in both cases and therefore the result is almost insensitive to the level of the noise. For the one-loop power spectrum the fact that the constraints are similar for S3 and S4 experiments is somewhat of a coincidence, due to the relative sizes of the theoretical error and the noise. When the error and the noise are combined, they are not very different for two different types of experiment in the range of $l$ where the most of the signal is coming from.

In Tab.~\ref{tab:S3-S4} we give unmarginalized constraints and results for an ideal case with no marginalization and no theoretical error. In the ideal case, the constraints from S3 and S4 type of experiments are different by a factor of 2. 

\begin{table}
\begin{center}
\begin{tabular}{c c c c}
\hline
{-} &  {\bf Lin$_{\rm non\;marg.}$} & {\bf 1L$_{\rm non\;marg.}$} & {\bf Ideal} \\
\hline
{\bf S3$_{\epsilon=0.05}$}\; & $35\; {\rm meV}\;\;$ & $17\; {\rm meV}\;\;$ & $15\; {\rm meV}\;\;$ \\
{\bf S3$_{\epsilon=0.10}$}\; & $46\; {\rm meV}\;\;$ & $20\; {\rm meV}\;\;$ & $15\; {\rm meV}\;\;$ \\
{\bf S4$_{\epsilon=0.05}$}\; & $31\; {\rm meV}\;\;$ & $14\; {\rm meV}\;\;$ & $8\; {\rm meV}\;\;$\\
{\bf S4$_{\epsilon=0.10}$}\; & $43\; {\rm meV}\;\;$ & $17\; {\rm meV}\;\;$ & $8\; {\rm meV}\;\;$\\
\hline
\end{tabular}
\caption{$1\sigma$ errors for the fiducial sum of neutrino masses $M_{\nu}=60 \; {\rm meV}$ for a S4 like experiment.}
\label{tab:S3-S4}
\end{center}
\end{table}

}

{\section{Conclusions}
\noindent  In this paper we showed that a consistent implementation of theoretical errors significantly modifies the constraints on some cosmological parameters. The reason is that the typical size of the theoretical errors is $\mathcal O(1\%)$ and this can be significantly larger than the signal of interest. We showed that constraints on neutrino mass and equilateral NG are worse by a factor of few once the theoretical errors are included. This will make measurements of these quantities more challenging than naively expected. Still, using a joint power spectrum and bispectrum analysis the significant detection of the minimal neutrino mass is possible by mapping galaxies up to $z\approx 2$. The prospects of measuring the neutrino mass using the CMB lensing are also good. For future polarization experiments the theoretical error can be made negligibly small using the two-loop matter power spectrum. On the other hand, improving the CMB limits on equilateral NG will be much more challenging. Our analysis implies that futuristic galaxy surveys can potentially reach $\sigma(f_{\rm NL}^{\rm eq.})\sim 10$ with current theoretical uncertainties. Local NG has a shape very orthogonal to the gravitational contributions, and for that reason the theoretical errors do not degrade the constraints significantly. Even at relatively low redshifts it seems possible to reach $\sigma(f_{\rm NL}^{\rm loc.})\sim 1$. 

Our results are relevant to making forecasts for future galaxy surveys but also to analysis of data sets with small statistical errors. We focused on two interesting examples of neutrino mass and primordial NG, but the method we propose applies more generally to any observable. We used a number of assumptions which do not affect the general features but can slightly change the numbers we quote. The most relevant assumptions are about the magnitude, the shape and the coherence length of the theoretical error. Different choices can lead to slightly different results. These properties are a priori unknown and one interesting question that requires further investigation is how to get reliable estimates for the error using the EFT of LSS, particularly for biased tracers. On the theory side, to make more precise forecasts, one also has to calculate the full one-loop bispectrum including primordial NG and all relevant bias parameters. First steps towards this goal were made in \cite{Baldauf:2014qfa, Angulo:2014tfa, Assassi:2015jqa,Angulo:2015eqa}. 

Another set of assumptions we made is about observable tracers and their bias parameters. We used only one bias parameter for all the redshift bins which is unrealistic and leads to more optimistic constraints. The results also depend on the choice of the shot noise and its redshift dependence. We used somewhat optimistic numbers that in each particular forecast have to be replaced with realistic survey/tracer dependent quantities.

}

\vskip.3cm

\noindent 
\emph{Acknowledgements.---}We would like to thank Daniel Baumann, Neal Dalal, Daniel Green and Anze Slosar for useful discussions and Enrico Pajer and Yvette Welling for helpful comments on the manuscript. T.B. gratefully acknowledges support from the Association of Members of the Institute for Advanced Study and NSF grant PHY-0855425. M.M. is supported by NSF Grants PHY-1314311 and PHY-1521097. M.S. gratefully acknowledges support from the Institute for Advanced Study and the Raymond and Beverly Sackler Foundation. M.Z. is supported in part by the NSF grants PHY-1213563, AST-1409709 and PHY-1521097.

\appendix
\section{Discrete Tracers}
In this Appendix we will rederive the stochasticity contributions to the power spectrum and bispectrum.
Let us consider a finite number of tracers $N$ such as galaxies at positions $\vec x_i$ in a finite volume $V$.
Their Fourier space density field (for $\vec k\neq 0$) is then given by
\be
\delta_g(\vec k)=\frac{1}{n}\sum_i \exp\left[i \vec k \vec x_i\right]\ ,
\ee
where $n=N/V$.
The power spectrum of the discrete tracers in the finite volume can then be computed as
\be
\begin{split}
&P_{g}=\frac{1}{V}\left\langle \delta_g(\vec k) \delta_g(-\vec k)\right\rangle\; ,\\
\;=&\frac{V}{N^2} \left[ \sum_{i=j}\left\langle\exp\left[i \vec k (\vec x_i-\vec x_j)\right]\right\rangle + \sum_{i\neq j}\left\langle\exp\left[i \vec k (\vec x_i-\vec x_j)\right]\right\rangle \right] \\
\;=&\frac{1}{n}+P_{g,\text{cont}}(k)\; .
\end{split}
\ee
Here, the constant $1/n$ is denoted the shot noise term and we have identified the non-zero separation expectation value with the continuous part of the discrete tracer power spectrum $P_{g,\text{cont}}(k)$. In the local bias model at linear order we have $P_{g,\text{cont}}(k)=b_1^2 P(k)$. 
This is clearly just an approximation for the continuous part of the tracer correlation function. Indeed, ref. \cite{Baldauf:2013hka} has argued that halo exclusion will alter this term and lead to an effective reduced stochasticity on large scales. Further corrections arise from the loop corrections to the continuous tracer power spectrum, for instance from the $b_2^2 \int d^3q/(2\pi)^3 P^2(q)$ contribution. Since these corrections are not under perturbative control, we introduce a free parameter $s_p$ that accounts for deviations from the fiducial $1/n$
\be
P_g(k)=s_p+P_{g,\text{cont}}(k)\; .
\ee

Let us now consider the bispectrum
\be
B_g=\frac{1}{V}\left\langle \delta_g(\vec k_1) \delta_g(\vec k_2)\delta_g(-\vec k_1-\vec k_2)\right\rangle \; .
\ee
Following the same steps that lead to the power spectrum above, we obtain
\be
\begin{split}
B_g=&\frac{V^2}{N^3}\sum_{i=j=l}\left\langle\exp\left[i \vec k_1 (\vec x_i-\vec x_l)+i \vec k_2 (\vec x_j-\vec x_l)\right]\right\rangle\\
+&3\frac{V^2}{N^3}\sum_{i=l\neq j}\left\langle\exp\left[i \vec k_1 (\vec x_i-\vec x_l)+i \vec k_2 (\vec x_j-\vec x_l)\right]\right\rangle\\
+&\frac{V^2}{N^3}\sum_{i\neq j\neq l}\left\langle\exp\left[i \vec k_1 (\vec x_i-\vec x_l)+i \vec k_2 (\vec x_j-\vec x_l)\right]\right\rangle\; .
\end{split}
\ee
This sum can be rewritten as
\be
B_g=\frac{1}{{n}^2}+\frac{1}{n}\left[P_{g,\text{cont}}(k_1)+2 \;\text{perm.}\right]+B_{g,\text{cont}}\; .
\ee

Again, the non-zero separation correlators are identified with the continuous power spectrum and bispectrum of the tracer field. We see that two different stochasticity corrections arise: a $1/{n}^2$ constant shot noise term and a product of the shot noise and the continuous power spectrum. As for the power spectrum discussed above, exclusion effects will alter both of these terms separately. Further corrections arise from clustering: For the constant term, a $b_2^3\int d^3q/(2\pi)^3P^3(q)$ correction leads to an effective stochasticity, that depends on the small scale power spectrum and is not under perturbative control. To absorb this clustering correction and the exclusion corrections,  we introduce a first free stochastic variable $s_{b,1}$. The mixed term is renormalized by a term $b_2 b_3 \int d^3q/(2\pi)^3 P^2(q)$. We see that this correction is different from both the constant stochasticity correction in the bispectrum and in the power spectrum, motivating another free stochasticity term $s_{b,2}$. Finally, we have for the discrete tracer bispectrum
\be
B_\text{g}=s_{b,1}+s_{b,2}\left[P_\text{g,cont}(k_1)+2 \;\text{perm.}\right]+B_\text{g,cont}\; .
\ee
In their role as counterterms in the EFT, the stochasticity corrections would only be required in conjunction with the corresponding loop terms, i.e., only once the loop corrections to the galaxy power spectrum and bispectrum are considered. However, their $k^0$ scaling and the fact that non-perturbative exclusion corrections contribute to their amplitude, motivate us to consider these terms even at tree level.

\bibliography{fnleq_lss}

\end{document}